\documentclass[aps,prd,preprint,nofootinbib,superscriptaddress,showpacs]{revtex4}
\usepackage{graphicx}
\usepackage{amsmath}
\usepackage{amsfonts}
\usepackage{feynmf}
\usepackage{hyperref}
\usepackage{subfigure}

\def \ba {\mathbf{A}}

\def \lvac {\langle0\vert}
\def \rvac {\vert0\rangle}
\def \br {\mathbf{r}}
\def \brg {\mathbf{R}}

\def \bx {\mathbf{x}}
\def \by {\mathbf{y}}
\def \bq {\mathbf{q}}

\def \cf {C_F}
\def \nc {N}

\def \nf {n_f}

\def \bk {\mathbf{k}}

\def \cc {\mathcal{C}}

\def\lQ{\Lambda_{\rm QCD}}

\def\als{\alpha_{\rm s}}
\newcommand{\MS}{\overline{\rm MS}}
\def\siml{{\ \lower-1.2pt\vbox{\hbox{\rlap{$<$}\lower6pt\vbox{\hbox{$\sim$}}}}\ }} 
\def\simg{{\ \lower-1.2pt\vbox{\hbox{\rlap{$>$}\lower6pt\vbox{\hbox{$\sim$}}}}\ }}

\def\vbfD{{\ \lower-8pt\vbox{\hbox{\rlap{$\!\leftrightarrow$}\lower8pt\vbox{\hbox{$\!\bf D$}}}}\ }} 
\def\dsl{\,\raise.15ex\hbox{/}\mkern-13.5mu D}

\newcommand{\nn}{\nonumber}

\newlength{\fillupPatternLength} 
\newlength{\fillupTextLength}

\newcommand{\Appendix}[1]%
    {%
     \section{#1}%
      }

\begin{document}
\setlength{\unitlength}{1mm}
\preprint{IFUM-915-FT \quad TUM-EFT 1/09}
\title{The three-quark static potential in  perturbation theory}
\author{Nora Brambilla}
\affiliation{Physik-Department, Technische Universit\"at M\"unchen,
James-Franck-Str. 1, 85748 Garching, Germany}
\author{Jacopo Ghiglieri}
\affiliation{Physik-Department, Technische Universit\"at M\"unchen,
James-Franck-Str. 1, 85748 Garching, Germany}
\affiliation{Excellence Cluster Universe, Technische Universit\"at M\"unchen, 
Boltzmannstr. 2, D-85748, Garching, Germany}
\author{Antonio Vairo}
\affiliation{Physik-Department, Technische Universit\"at M\"unchen,
James-Franck-Str. 1, 85748 Garching, Germany}


\begin{abstract}
We study the three-quark static potential in perturbation theory in QCD.
A complete next-to-leading order calculation is performed in the singlet,
octets and decuplet channels and the potential exponentiation is demonstrated.
The mixing of the octet representations is calculated.
At next-to-next-to-leading order, the subset of diagrams producing three-body forces is 
identified in Coulomb gauge and its contribution to the potential calculated.
Combining it with the contribution of the two-body forces, which may be extracted 
from the quark-antiquark static potential, we obtain the complete next-to-next-to-leading order 
three-quark static potential in the colour-singlet channel. 
\end{abstract}

\pacs{12.38.-t,12.38.Bx,14.20.-c,14.40.Pq}

\maketitle

\section{Introduction}  
\label{sec:introd}
The interaction among heavy quarks has been explored since the
QCD inception as an important tool to learn about the characteristics
of the non-Abelian gauge dynamics in general and the QCD low-energy behaviour
in particular \cite{Wilson:1974sk,Susskind:1976pi}.  

The $Q \bar{Q}$ static potential is a very well known quantity 
for its crucial role in quarkonium phenomenology \cite{Brambilla:2004wf,Godfrey:2008nc}
and for having been studied extensively by lattice gauge 
theories since their introduction \cite{Bali:2000gf}. The typical shape of the colour-singlet 
$Q \bar{Q}$ static potential, which is characterized by a short-range Coulomb 
behaviour and a long-range linear rise, well represents the double nature of 
QCD as an asymptotically free and infrared confined theory.
Also gluonic excitations of a static quark-antiquark pair 
have been explored by lattice calculations both in the long range, 
where they exhibit a stringy behaviour like the 
colour-singlet potential, and in the short range where 
they show a Coulomb-like behaviour in one of the two possible 
quark-antiquark colour configurations: singlet or octet  
\cite{Bali:2000gf,Juge:2002br}.

More recently, non-relativistic effective field theories 
of QCD have provided a new way to look at the quark-antiquark 
potential and allowed, specially in the short range, calculations 
with unprecedented precision \cite{Brambilla:2004jw}.
Presently, the static quark-antiquark potential is completely known 
up to two loops \cite{Peter:1997me,Schroder:1998vy}. Starting from 
three loops the potential exhibits infrared divergences;  these have been 
calculated at leading order (LO) \cite{Brambilla:1999qa} and next-to-leading 
order (NLO) \cite{Brambilla:2006wp}, and resummed at 
leading logarithmic (LL) \cite{Pineda:2000gza} and next-to-leading 
logarithmic (NLL) \cite{Brambilla:2009bi} accuracy. The fermionic part 
of the three-loop finite contribution has been calculated 
recently \cite{Smirnov:2008pn}. High-order perturbative 
calculations show a remarkably good agreement  
with the lattice determinations of the static quark-antiquark 
energy up to a distance of about 0.2 - 0.3 fm 
\cite{Sumino:2001eh,Necco:2001gh,Pineda:2002se,Brambilla:2009bi}, 
which allows to constrain the size of the unknown higher-order contributions.

The static quark-antiquark energy may be extracted from the 
large-time behaviour of the static quark-antiquark Wilson loop. 
Extremely accurate lattice determinations of the static energy 
at short distances (the smallest distance being about 0.08 fm) 
can be found in \cite{Necco:2001xg}.
Also gluonic excitations between static quark-antiquark sources 
have been explored in the framework of effective field theories \cite{Brambilla:1999xf} 
and by means of lattice calculations \cite{Juge:2002br}.
Again, high-order perturbative calculations show agreement with 
accurate short-range lattice data and allow for the precise extraction of 
the so-called gluelump masses \cite{Bali:2003jq}. 
Short distance studies of the quark-antiquark interaction tell us about the interplay of perturbative 
and non-perturbative contributions in QCD, in particular that perturbative contributions describe the data 
with a high accuracy up to distances of 0.2 - 0.3 fm, while 
a confining string sets in only at distances of about 0.5 fm  \cite{Luscher:2002qv}, 
and that the operator product expansion does not appear to be violated. 
It is only natural to ask if these features are specific of quark-antiquark systems, 
i.e. mesons, or may also show up, and, in case, to which extent, 
in three-quark systems, i.e. baryons.

The potential that describes the interaction of three heavy quarks $Q$ 
is much less known than the heavy $Q\bar{Q}$ potential, 
one of the reasons  being the difficulty of producing 
$QQQ$ states and the consequent lack of experimental data.  
This has led to a wide use of phenomenolo\-gical models \cite{Richard:1992uk,Klempt:2009pi}, 
sometimes based on strong-coupling expansion arguments and lattice evaluations
of the three-quark static Wilson loop; often a sum of two-body interactions has been used.

A rigorous definition of the $QQQ$ potential is provided by the 
non-relativistic effective field theory for $QQQ$ states formulated 
in \cite{Brambilla:2005yk} ($QQq$ states have been considered in \cite{Brambilla:2005yk,Fleming:2005pd}).
This effective field theory is the three heavy-quark version of potential 
non-relativistic QCD (pNRQCD), the effective field theory first introduced 
for quarkonium in \cite{Pineda:1997bj,Brambilla:1999xf}.
pNRQCD is constructed from QCD as an expansion in the inverse of the 
heavy-quark mass $m$ and in the distances between the heavy quarks 
(multipole expansion). At zeroth order in the multipole expansion, 
the equation of motion of pNRQCD is the Schr\"odinger equation with 
the potentials given by the Wilson coefficients of the six-fermion operators.
The Wilson coefficients are calculated by equating, i.e. matching, 
amplitudes in QCD with amplitudes in pNRQCD order by order in 
$1/m$ and in the multipole expansion. In particular, 
the static potentials of the different colour representations 
are evaluated by matching to static Wilson loops in QCD.
At distances shorter than the inverse of the typical hadronic scale, $\lQ$,  
the degrees of freedom of pNRQCD are a $QQQ$ colour-singlet field, 
two $QQQ$ colour-octet fields, a $QQQ$ colour-decuplet field, light quarks 
and low-energy gluons. The Wilson coefficients of the corresponding six-fermion 
operators are the singlet, octet and decuplet potentials respectively.
They may be evaluated in perturbation theory.
To the best of our knowledge only the LO expressions 
(excluding octet mixing) have been considered so far.
At distances larger than $1/\lQ$, when confinement sets in, the degrees of freedom of pNRQCD are
only the $QQQ$ colour-singlet field and light hadrons. 
Gluonic excitations of heavy-quark bound states cannot be resolved at 
such distances because of the mass gap of order $\lQ$ that they develop 
with respect to the colour-singlet state 
(cf. with the lattice data in \cite{Takahashi:2002it,Takahashi:2004rw}).  
In this situation, the matching to pNRQCD cannot be performed 
in perturbation theory but must rely on non-perturbative methods.
The non-perturbative static, spin-dependent and $1/m$  
colour-singlet $QQQ$ potentials have been expressed in terms of Wilson loops 
in \cite{Brambilla:2005yk} (for earlier work see \cite{Brambilla:1993zw,Brambilla:1995px}). 
So far only the static potential has been evaluated on the lattice.
 
Most of the existing lattice studies of the three-quark static potential
have explored the region of large interquark distances
\cite{Sommer:1985da,Bornyakov:2004uv,Bornyakov:2004yg,Alexandrou:2002sn,
Takahashi:2003ty,Suganuma:2000bi,Takahashi:2000te,Takahashi:2002bw,Takahashi:2004rw}.
As for the $Q\bar{Q}$ case, the characteristic signature  of the 
long-range non-Abelian dynamics  is believed to be 
a linear  ``stringy''  rising of the static interaction. Moreover, 
the general expectation for the baryonic case is that, at least
classically, the strings meet at the so called Fermat (or Torricelli) point, which has minimum 
distance from the three sources ($Y$-shape configuration).
If this is the case, one should see a genuine three-body interaction among the
static quarks. In another model \cite{Cornwall:1996xr},  the  long range  $QQQ$ potential  is simply
the sum of two-body potentials ($\Delta$-shape configuration). Most of the lattice
calculations of the $QQQ$ static potential have focused on
distinguishing the $Y$ configuration (favoured by data) from the $\Delta$ configuration, 
despite the difference between a $\Delta$ and a $Y$
shape potential being rather small  and difficult to detect.
Recently, however, some data have accumulated that include short 
distances both at zero and finite temperature, and both for the lowest and for some 
higher gluonic excitations \cite{Takahashi:2002it,Takahashi:2004rw,Hubner:2007qh}. 
This opens the possibility to address, also for the $QQQ$ system, 
questions about the short-range behaviour of the static potential and its gluonic excitations, 
and more specifically about the region of validity of perturbation theory  
and about the cross-over region from perturbative to non-perturbative QCD.
In general, one expects this cross-over to happen in a more spectacular 
way than in the quark-antiquark case, due to the overcoming of 
the long-range three-body forces over the short-range two-body Coulomb forces.

In the paper, we focus on the potential between three static quarks 
in the different colour configurations and at short distances.
Surprisingly, very little is known about it besides the LO expression.
For all colour configurations, we will perform a complete NLO calculation showing explicitly how the 
exponentiation works at this order. For the singlet and decuplet potentials, 
we will prove that the naive extension of the NLO two-body 
result turns out to be correct. For the octet potentials, 
we will need to account for the mixing, which already sets in at LO. 
At next-to-next-to leading order (NNLO) the first genuine three-body contribution 
appears. We calculate it for the singlet and decuplet colour configuration.
In the colour-singlet case, combining the three-body contribution with 
the two-body one that can be extracted from the quark-antiquark static potential,
we will obtain the complete NNLO potential.

The plan of the paper is the following.  
In Sec.~\ref{sec:threeqpot}, we introduce the three-quark Wilson loop and define the potential.
In Sec.~\ref{sec:tree},  we derive its expression at order $g^2$ 
for the singlet, the octets and the decuplet representations, 
showing  that the two octet representations mix.  
In Sec.~\ref{sec:one_loop}, we calculate the static potentials at order $g^4$ 
and show how exponentiation works at this order; a generalization of this 
result to $N$ quarks in SU($\nc$) is provided in Sec.~\ref{sec:N}. 
In Sec.~\ref{sec:3qg6}, we identify the first genuine three-body contribution 
to the potential that appears in perturbation theory  at order $g^6$ and 
evaluate it in several geometrical configurations. 
In Sec.~\ref{sec:nnlo}, we derive the two-body colour-singlet contribution and 
hence provide the complete colour-singlet static potential at order $g^6$.
Sec.~\ref{sec:conc} is devoted to the conclusions and a short outlook.  
Some technical details may be found in the appendices.

\section{The three-quark static potential}
\label{sec:threeqpot}
In this section, we consider the perturbative static potential 
of three heavy quarks. In the effective field theory language of \cite{Brambilla:2005yk}, 
the potentials in the different colour representations are the matching coefficients 
of the six-fermion operators made of two singlet, two octet or two decuplet fields.
The matching coefficients can be ordered in powers of $1/m$, 
the static potential corresponding to the first term in the series.
The perturbative expression of the potential is expected to describe correctly  
the potential at short distances $r$, for which $\als(1/r) \ll 1$ holds.

\begin{figure}
\begin{center}
\includegraphics{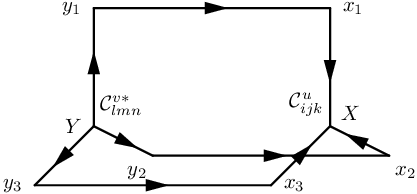}
\end{center}
\caption{\label{fig:wloop}
Static Wilson loop with edges 
$x_1 = ({\boldsymbol{x}}_1,T_W/2)$, $x_2 = ({\boldsymbol{x}}_2,T_W/2)$, $x_3 = ({\boldsymbol{x}}_3,T_W/2)$, 
$y_1 = ({\boldsymbol{x}}_1,-T_W/2)$, $y_2 = ({\boldsymbol{x}}_2,-T_W/2)$, $y_3 = ({\boldsymbol{x}}_3,-T_W/2)$ 
and insertions of the tensors  $\mathcal{C}^u_{ijk}$ and $\mathcal{C}^{v\,*}_{lmn}$
in $X = (\boldsymbol{R}, T_W/2)$ and $Y = (\boldsymbol{R},-T_W/2)$ respectively.}
\end{figure}

The static potential is computed by matching  the appropriate 
Green's function in QCD with static sources (Wilson loop) to the 
corresponding Green's function in pNRQCD \cite{Brambilla:1999xf,Brambilla:2004jw,Brambilla:2005yk}.
The Green's function in pNRQCD describes the propagation of a 
static $QQQ$ state in the colour representation $\cc$ through a potential $V_\cc$.
Loop corrections due to gluons of energy and momentum of order $\als/r$ contribute 
at  next-to-next-to-next-to-leading order (N$^3$LO) and are beyond the accuracy of this work.
The matching condition valid up to and including NNLO is, for $T_W \to \infty$,
\begin{equation}
\label{defpotential}
\lvac \cc^u W \cc^{v\dagger}\rvac =
Z_{\cc}(\mathfrak{r})\exp{(-iV_\cc(\mathfrak{r})T_W)}\lvac S^{uv}_\cc(T_W/2,-T_W/2)\rvac.
\end{equation}
The left-hand side stands for the expectation value of the 
three-quark static Wilson loop: a possible choice is shown in Fig.~\ref{fig:wloop}.
The static quarks are located in ${\boldsymbol{x}}_1$, ${\boldsymbol{x}}_2$, 
and ${\boldsymbol{x}}_3$ and propagate from the initial time $-T_W/2$ to the final time $T_W/2$.
The colour tensors  $\mathcal{C}^u$ and $\mathcal{C}^{v\,\dagger}$ are inserted in the Wilson 
loop in the centre-of-mass coordinate 
$\boldsymbol{R} = ({\boldsymbol{x}}_1 + {\boldsymbol{x}}_2 + {\boldsymbol{x}}_3)/3$ 
at the final and initial times respectively. $\rvac$ on the left-hand (right-hand) side stands 
for the vacuum state of QCD (pNRQCD). In the right-hand side,
$V_\cc$ stands for the static potential in the colour representation $\cc$, 
$Z_{\cc}$ for a normalization factor, $S_\cc$ for the Wilson loop $\cc^u W \cc^{v\dagger}$ with 
all the quarks located in the centre of mass, and $\mathfrak{r}=\{\br_{1},\br_{2},\br_{3}\}$ 
for the set of distances between the quarks, defined as
\begin{equation}
\label{defdistances}
\br_{1}=\bx_1-\bx_2,\qquad\br_2=\bx_1-\bx_3,\qquad\br_3=\bx_2-\bx_3;
\end{equation}
only two of these three distances are independent: $\br_{1}+\br_3 =\br_2$.
The explicit expressions of the three-quark static Wilson loop shown in Fig.~\ref{fig:wloop} 
and of $S^{uv}_\cc$ are  
\begin{eqnarray}
\nonumber 
\cc^u\,W\,\cc^{v\dagger} &=&
\mathcal{C}^u_{ijk}
\phi_{ii'}(\brg,\bx_1,T_W/2)\phi_{i'r}(T_W/2,-T_W/2,\bx_1)\phi_{rl}(\bx_1,\brg,-T_W/2)
\\
\nonumber 
&\times&\phi_{jj'}(\brg,\bx_2,T_W/2)\phi_{j's}(T_W/2,-T_W/2,\bx_2)\phi_{sm}(\bx_2,\brg,-T_W/2)
\\
&\times&\phi_{kk'}(\brg,\bx_3,T_W/2)\phi_{k't}(T_W/2,-T_W/2,\bx_3)\phi_{tn}(\bx_3,\brg,-T_W/2)\mathcal{C}^{v\dagger}_{lmn},
\label{3quarkwilson}
\end{eqnarray}
\begin{equation}
S^{uv}_\cc = \cc^u_{ijk}\phi_{il}(T_W/2,-T_W/2,\brg)\phi_{jm}(T_W/2,-T_W/2,\brg)\phi_{kn}(T_W/2,-T_W/2,\brg)\cc^{v\dagger}_{lmn}, 
\label{string3q}
\end{equation}
where repeated indices are implicitly summed from $1$ to $3$.
The tensor $\cc^u$ is inserted at $X=({\bf R}, T_W/2)$, 
while its conjugate $\cc^{v\dagger}$ is inserted at $Y=({\bf R}, -T_W/2)$. 
The function $\phi$ stands for a Wilson line: the spacelike Wilson line at time $t$ reads
\begin{equation}
\label{spacewilsonline}
\phi(\by,\bx,t) = {\rm P}\,\exp\left(ig\int_0^1ds\,(\by-\bx)\cdot\ba(\bx+(\by-\bx)s,t)\right),
\end{equation}
while the timelike Wilson line at position ${\bf x}$ reads 
\begin{equation}
\label{timewilsonline}
\phi(t_f,t_i,\bx) = {\rm P}\,\exp\left(ig\int_{t_i}^{t_f}dt\,A^0(t,\bx)\right).
\end{equation}
In both expressions,  $A_\mu=A_\mu^aT^a$ and  ${\rm P}$ stands for the path ordering of the 
matrices $A_\mu$ along the Wilson line.
In Eqs. (\ref{3quarkwilson}) and (\ref{string3q}), we have explicitly written the colour indices 
of the Wilson lines in the fundamental representation. 

Let us now specify the colour representations $\cc^u$. 
A $QQQ$ state can be decomposed into the following representations: 
\begin{equation}
\label{qqqtensor}
3\otimes3\otimes3=1\oplus8\oplus8\oplus10 ,
\end{equation}
where the singlet representation is totally antisymmetric, 
the decuplet is totally symmetric, and the two octets have mixed symmetries. 
A generic representation $\cc^u$ has three  colour indices, $i,j,k$,  
running from $1$ to $3$ and is written in detail as $\mathcal{C}^u_{ijk}$. 
The labels $u,v$ refer to the type of colour representations, specifically,  
when $\cc$ and $\cc^\dagger$ are both in the singlet representation, the indices $u$ and $v$ are suppressed; 
when $\cc^u$ and $\cc^{v\dagger}$ are both in the decuplet representation, $u$ and $v$ range from $1$ to $10$;
when $\cc^u$ and $\cc^{v\dagger}$ are in the antisymmetric or in the symmetric octet representations, the 
indices $u$ and $v$ range from $1$ to $8$. 
The concrete choice that we have operated for these rank-three tensors is given in  Appendix \ref{app_rep}.
In the singlet and decuplet cases, $\cc^u$ and $\cc^{v\dagger}$ are real numbers.
In the octet case, since the octets mix, it is more convenient 
to consider $\cc^u$ and $\cc^{v\dagger}$ as 2 component vectors; we will detail about this in the next section.

The quantity $\lvac S^{uv}_\cc(T_W/2,-T_W/2)\rvac$ is dimensionless. In perturbation theory, 
it may depend on $T_W$ only logarithmically, therefore  
$\displaystyle \lim_{T_W\to\infty} 1/T_W \times \ln \lvac S^{uv}_\cc(T_W/2,-T_W/2)\rvac = 0$. 
Also  $\displaystyle \lim_{T_W\to\infty} 1/T_W \times \ln Z_{\cc}(\mathfrak{r}) =0$, 
because $ Z_{\cc}(\mathfrak{r})$ does not depend on $T_W$.
Hence, the matching condition (\ref{defpotential}) may be rewritten as 
\begin{equation}
\label{defpotential2}
V_\cc(\mathfrak{r})=\lim_{T_W\to\infty} 
\frac{i}{T_W}\ln \frac{\lvac \cc^u\,W\,\cc^{v\dagger}\rvac}{
\cc^u_{mno}{\cc}^{v\dagger}_{mno}} ,
\end{equation}
where we have kept in the denominator a colour tensor normalization factor (cf. Eq. \eqref{normortqqq}).
It is convenient to define 
\begin{equation}
\label{expansionW}
\frac{\lvac \cc^u\,W\,\cc^{v\dagger}\rvac}{
\cc^u_{mno}{\cc}^{v\dagger}_{mno}} =   
1+ \mathcal{M}^{(0)}(\cc,\mathfrak{r}) + \mathcal{M}^{(1)}(\cc,\mathfrak{r}) 
+ \mathcal{M}^{(2)}(\cc,\mathfrak{r}) + \dots \,,
\end{equation}
with the quantities $\mathcal{M}^{(n)}$ encoding all contributions of order $g^{2n+2}\sim \als^{n+1}$
for a given colour representation $\cc$. 
Analogously we may write 
\begin{equation}
V_\cc(\mathfrak{r}) = V^{(0)}_\cc(\mathfrak{r}) + V^{(1)}_\cc(\mathfrak{r})+V^{(2)}_\cc(\mathfrak{r}) + \dots\,,
\label{PTpotential}
\end{equation} 
where $ V^{(n)}_\cc(\mathfrak{r})$ encodes all contributions of order $g^{2n+2}$ to the potential.
From Eqs. (\ref{defpotential2}), (\ref{expansionW}) and (\ref{PTpotential}), the order by order  
matching conditions for the potential read 
\begin{eqnarray}
\label{PTpotential0}
V_\cc^{(0)}(\mathfrak{r})&=&\lim_{T_W\to\infty} \frac{i}{T_W}  \mathcal{M}^{(0)}(\cc,\mathfrak{r}) , 
\\
\label{PTpotential1}
V_\cc^{(1)}(\mathfrak{r})&=&\lim_{T_W\to\infty} \frac{i}{T_W}  
\left(\mathcal{M}^{(1)}(\cc,\mathfrak{r}) - \frac{1}{2}\mathcal{M}^{(0)\,2}(\cc,\mathfrak{r})\right), 
\\
\label{PTpotential2}
V_\cc^{(2)}(\mathfrak{r})&=&\lim_{T_W\to\infty} \frac{i}{T_W}  
\left(\mathcal{M}^{(2)}(\cc,\mathfrak{r}) - \mathcal{M}^{(0)}(\cc,\mathfrak{r}) \mathcal{M}^{(1)}(\cc,\mathfrak{r})  
+ \frac{1}{3}\mathcal{M}^{(0)\,3}(\cc,\mathfrak{r})\right), 
\\
\nn
\cdots  & & \qquad \cdots \qquad.
\end{eqnarray}
Note that the subtraction terms, $\mathcal{M}^{(0)\,2} \sim T_W^2$, 
$\mathcal{M}^{(0)}\mathcal{M}^{(1)} \sim T_W^2$ and  $\mathcal{M}^{(0)\,3} \sim T_W^3$, are 
divergent in the $T_W \to \infty$ limit. They cancel against divergences in 
$\mathcal{M}^{(1)}$ and $\mathcal{M}^{(2)}$. Canceling the divergences may be  
interpreted as reconstructing the exponential $\exp{(-iV_\cc(\mathfrak{r})T_W)}$
in the matching condition (\ref{defpotential}). For this reason, the procedure 
of verifying the finiteness of the limits (\ref{PTpotential1}), (\ref{PTpotential2}), ... 
is  often referred to as verifying the \emph{potential exponentiation}.

\section{The static potential at LO}
\label{sec:tree} 
To set up the notation and to discuss the octet mixing, we start by calculating 
the three-quark static potential at LO, i.e. $V_\cc^{(0)}$.
The calculation can be split into two steps: the computation of the amplitudes and the 
calculation of the colour factors, which will differ for each potential. 
Throughout the paper we choose the Coulomb gauge for the calculation of the amplitudes, 
since it consistently reduces the number of diagrams to be computed. Of course, the calculated LO, NLO and NNLO potentials are gauge invariant\footnote{ 
Possible complications arising in Coulomb gauge because of the so-called 
Schwinger--Christ--Lee terms \cite{Schwinger:1962wd,Christ:1980ku,Kummer:1994bq}, 
which involve $\als^2$-suppressed non-local interactions with transverse gluons, 
affect the potential at next-to-next-to-next-to leading order or smaller 
and are beyond the accuracy of the present work. 
}.

\begin{figure}
\begin{center}
\includegraphics[scale=1]{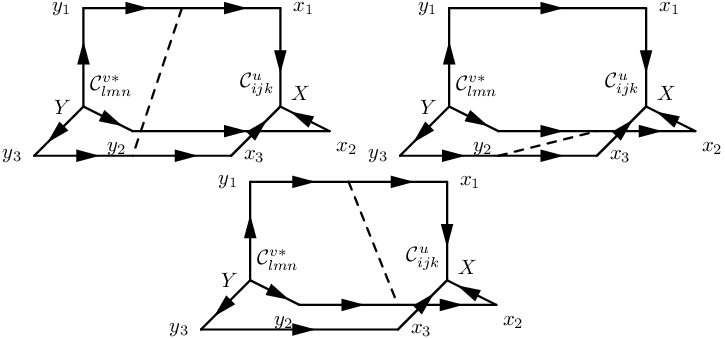}
\end{center}
\caption{The three terms contributing to the static potential at order $g^2$. Dashed lines are longitudinal gluons.}
\label{fig:wloop3one}
\end{figure}

\begin{figure}
\begin{center}
\includegraphics{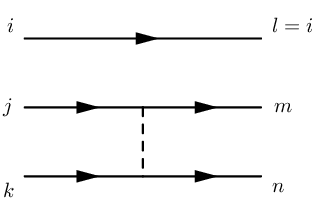}
\end{center}
\caption{A diagram contributing to the mixing of the two octets at LO. The three quark lines 
represent from above to below the quarks in ${\bf x}_1$, ${\bf x}_2$ and ${\bf x}_3$ respectively.}
\label{fig:oaos}
\end{figure}

At order $g^2$, the diagrams  that contribute to the potential
are those shown in Fig.~\ref{fig:wloop3one}, in which a gluon is
exchanged between two quark lines, thus leaving the third quark line
untouched. In the following, we will call such a line a \emph{spectator line}. 
Since diagrams involving gluon exchanges between quark lines and strings 
contribute  only to the normalization $Z_\cc$,\footnote{
Our work is concerned with the static potential up to NNLO.
Up to this order, two-body diagrams involving gluon exchanges with strings are 
of the same type as those encountered in the evaluation of the quark-antiquark 
potential and do not contribute to the potential by the same arguments 
used there \cite{Fischler:1977yf,Peter:1997me,Schroder:1998vy} (a more detailed discussion can be found 
in \cite{Schroder:1999sg}). 
At NNLO order, there is also a class of three-body diagrams involving 
gluon exchanges with the strings. This class of diagrams has a transverse gluon
emitted from one string and three longitudinal gluons coupled to it and to three 
different quark lines. These diagrams vanish because either they involve triple-gluon vertices 
with two transverse and one longitudinal gluon, but zero inflowing energy 
(taking $T_W\to\infty$ is equivalent to set to zero the energy flowing from the quark lines, 
see footnote \ref{footq0}), or they involve quartic-gluon 
vertices with one transverse and three longitudinal gluons.
}
we will adopt, in the following, a simpler representation of the diagrams 
without end-point strings (see e.g. Fig.~\ref{fig:oaos}). It is convenient to define
$\displaystyle \mathcal{M}^{(0)}(\cc,\mathfrak{r}) = \sum_{q=1}^3\mathcal{M}_{q}^{(0)}(\cc,\br_{q})$, 
where $\mathcal{M}_{q}^{(0)}(\cc,\br_{q})$ is the amplitude of the one-gluon
exchange between two of the three sources: $q=1$ corresponds to the exchange between the quark 
in $\bx_1$ and the one in $\bx_2$, $q=2$  corresponds to the exchange between the quark in $\bx_1$ 
and the one in $\bx_3$, and $q=3$  corresponds to the exchange between the quark in $\bx_2$ 
and the one in $\bx_3$.\footnote{For a baryon of $N$ quarks in SU$(\nc)$, 
there will be $\nc(\nc-1)/2$ possible gluon exchanges, with $(\nc-2)$ spectator quarks.} 
The  potential at order $g^2$ then reads 
\begin{equation}
\label{defpotentialwm}
V^{(0)}_\cc(\mathfrak{r})=\lim_{T_W\to\infty} \frac{i}{T_W}\sum_{q=1}^3\mathcal{M}_{q}^{(0)}(\cc,\br_{q})
=\sum_{q=1}^3f^{(0)}_q(\cc)\frac{\alpha_s}{\vert\br_q\vert}.
\end{equation}

The colour part of the amplitude has been factored in the colour coefficient $f^{(0)}_q(\cc)$. 
This coefficient is defined as
\begin{equation}
\label{colourfactortreelevel}
f^{(0)}_q(\cc)=\frac{
\cc^u_{jkl} {\cal T}^{q\,(0)}_{jj'kk'll'}
\cc^{v\dagger}_{j'k'l'}}{
\cc^u_{mno}{\cc}^{v\dagger}_{mno}},
\end{equation} 
where ${\cal T}^{1\,(0)}_{jj'kk'll'} = T^a_{jj'} T^a_{kk'}\delta_{ll'}$, 
${\cal T}^{2\,(0)}_{jj'kk'll'} = T^a_{jj'} \delta_{kk'}T^a_{ll'}$ and 
${\cal T}^{3\,(0)}_{jj'kk'll'} = \delta_{jj'} T^a_{kk'}T^a_{ll'}$.

In the singlet case, $\cc= \cc^*= S$ (see Eq. \eqref{baryonsinglet}), we have 
\begin{equation}
\label{f0singlet}
f^{(0)}_1(S)=f^{(0)}_2(S)=f^{(0)}_3(S)=-\frac{2}{3}, 
\end{equation} 
and the LO  colour-singlet static potential has the well-known form:
\begin{equation}
\label{potsinglet3}
V^{(0)}_s(\mathfrak{r})=
-\frac{2}{3}\als\left(\frac{1}{\vert\br_{1}\vert}+\frac{1}{\vert\br_{2}\vert}+\frac{1}{\vert\br_{3}\vert}\right).
\end{equation}
We note that in the limit where one quark is put at infinite distance 
the above potential should reproduce one of the two quark-quark 
potentials, either the antisymmetric antitriplet one or the symmetric sextet one. 
Since the singlet is antisymmetric one recovers indeed 
the antisymmetric triplet quark-quark potential \cite{Flamm:1982jv}.
Moreover, we observe that the colour-singlet quark-antiquark potential is twice 
each quark-quark component of \eqref{potsinglet3}: we will generalize this result in Sec.~\ref{sec:N}.

In the decuplet case, $\cc^u= \cc^{u*}= \Delta^u$ (see Eq. \eqref{decuplet}), we have  
\begin{equation}
\label{f0decuplet}
f^{(0)}_1(\Delta)=f^{(0)}_2(\Delta)=f^{(0)}_3(\Delta)=\frac{1}{3},
\end{equation}
and the LO  colour-decuplet static potential reads:
\begin{equation}
\label{potdecuplet}
V^{(0)}_d(\mathfrak{r})=\frac{1}{3}\als
\left(\frac{1}{\vert\br_{1}\vert}+\frac{1}{\vert\br_{2}\vert}+\frac{1}{\vert\br_{3}\vert}\right).
\end{equation}
We note that in the limit where one quark is put at infinite distance the above potential reproduces 
the symmetric sextet quark-quark potential \cite{Flamm:1982jv}.

In the octet case, the one-gluon exchange mixes the symmetric 
and the antisymmetric octets, i.e. there is a nonzero colour amplitude with an initial 
symmetric octet state and a final antisymmetric one and viceversa.
It is, therefore, convenient to define a potential $V_O(\mathfrak{r})$, which is a $2\times2$ matrix, 
and a vector colour representation:  $\displaystyle \cc^a = O^a_{ijk}=\binom{O^{Aa}_{ijk}}{O^{Sa}_{ijk}}$.  
A possible choice for the symmetric and antisymmetric octet representations $O^{Sa}_{ijk}$ and  $O^{Aa}_{ijk}$ 
is in Eqs. \eqref{os} and \eqref{oa} respectively: 
in this choice, both representations are symmetric or antisymmetric in the first 
two indices $i$ and $j$. According to the definition  \eqref{3quarkwilson}, 
the third index, $k$, is associated to the quark in ${\bf x}_3$, therefore  
we expect that the diagrams responsible for the mixing are those 
involving gluons attached to the third quark line, like the one shown in Fig.~\ref{fig:oaos}. 
Indeed, by computing $\mathcal{M}_{q}^{(0)}(O,\br_{q})$ it follows that 
the three $2\times2$ matrices $f^{(0)}_q(O)$ are given by 
\begin{equation}
\label{f0octet}
f^{(0)}_1(O) =
\left(\begin{array}{cc}-\frac{2}{3}&0\\
0&\frac{1}{3}\end{array}\right), \quad
f^{(0)}_2(O) =
\left(\begin{array}{cc}\frac{1}{12}&-\frac{\sqrt{3}}{4}\\
-\frac{\sqrt{3}}{4}&-\frac{5}{12}\end{array}\right),\quad
f^{(0)}_3(O) =
\left(\begin{array}{cc}\frac{1}{12}&\frac{\sqrt{3}}{4}\\ 
\frac{\sqrt{3}}{4}&-\frac{5}{12}\end{array}\right),
\end{equation} 
where we see that $f^{(0)}_1(O)$ is diagonal but  $f^{(0)}_2(O)$ and $f^{(0)}_3(O)$ are not.
Hence, the LO colour-octet static potential reads 
\begin{equation}
\label{potoctet}
V^{(0)}_O(\mathfrak{r})=\als\left[\frac{1}{\vert\br_1\vert}
\left(\begin{array}{cc}-\frac{2}{3}&0\\
0&\frac{1}{3}\end{array}\right)+\frac{1}{\vert\br_2\vert}
\left(\begin{array}{cc}\frac{1}{12}&-\frac{\sqrt{3}}{4}\\
-\frac{\sqrt{3}}{4}&-\frac{5}{12}\end{array}\right)+\frac{1}{\vert\br_3\vert}
\left(\begin{array}{cc}\frac{1}{12}&\frac{\sqrt{3}}{4}\\ 
\frac{\sqrt{3}}{4}&-\frac{5}{12}\end{array}\right)\right].
\end{equation}
The part of the potential proportional to ${1}/{\vert\br_1\vert}$
is diagonal and its entries are equal to the ${1}/{\vert\br_1\vert}$ 
parts of the colour-singlet and colour-decuplet potentials. 
This can be explained by observing that if the quark in ${\bf x}_3$ 
is put to infinity  the two octets disentangle and we
are left with two, antisymmetric and symmetric, quark-quark potentials.
The parts of the potential proportional to  ${1}/{\vert\br_2\vert}$
and  ${1}/{\vert\br_3\vert}$ have the same diagonal
elements but opposite off-diagonal ones: this means that they share
the same eigenvalues but have different eigenvectors. The eigenvalues
are $(-2/3,1/3)$ with corresponding eigenvectors
\begin{equation}
\label{eigenoctet}
\lambda_{-2/3}=\left(\mp \frac{O^A}{\sqrt{3}}\;,\;O^S\right),\qquad
\lambda_{1/3}=\left(\pm\sqrt{3}O^A\;,\;O^S\right),
\end{equation} 
where the upper sign refers to the matrix with positive
off-diagonal elements and the lower sign to the other one. If we construct
a matrix $P$ such that $PMP^{-1}$ is diagonal, where $M$ is one
of the two non-diagonal matrices, the other being $M'$, then 
neither  $PM'P^{-1}$ nor $P f^{(0)}_1(O) P^{-1}$ are diagonal.
The diagonalization of the part of the potential proportional to the 
distance ${1}/{\vert\br_{2}\vert}$ (${1}/{\vert\br_{3}\vert}$)
thus simply corresponds to changing to a new octet 
representation symmetric and antisymmetric in the indices $i$ and $k$ ($j$ and $k$).
Note that by pulling at infinite distance the quark in ${\bf x}_2$ or in ${\bf x}_1$
we are left with a matrix, which, after diagonalization, reproduces again the 
two, antisymmetric and symmetric, quark-quark potentials.

The fact that the two octets mix has, to our knowledge, 
not been discussed in the literature so far. The octet 
$QQQ$  potential can be extracted from the lattice data in \cite{Hubner:2007qh}. 
There, equilateral geometries ($\vert\br_1\vert = \vert\br_2\vert
= \vert\br_3\vert$) have been taken into account for which the off-diagonal 
elements cancel (see Eq. \eqref{potoctet}).\footnote{For equilateral geometries, 
the singlet and octet potentials are attractive while the decuplet one is repulsive.  
} 
In general, off-diagonal elements cancel in any isosceles geometry.
Clearly, the mixing needs instead to be properly accounted for in any lattice 
simulation based on non-isosceles geometries.

\section{The static potential at NLO}
\label{sec:one_loop} 
\begin{figure}
\begin{center}
\includegraphics{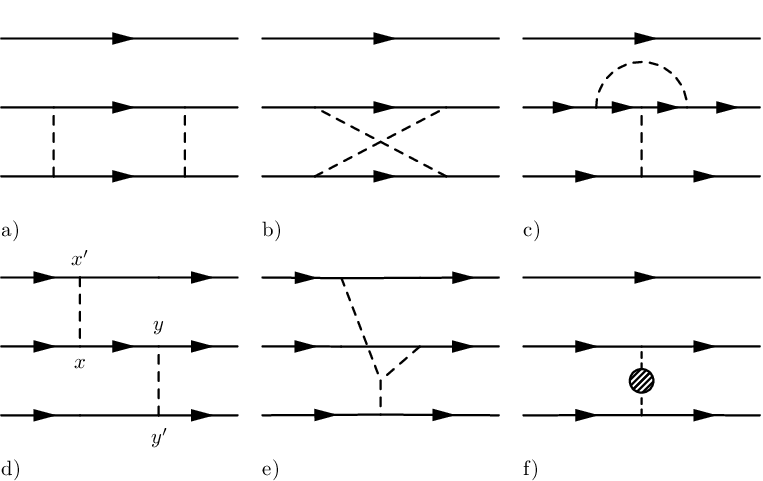}
\end{center}
\caption{Diagrams appearing at order $g^4$ in the three-quark potential.}
\label{fig:3quark1loop}
\end{figure}

The NLO, i.e. the order $g^4$, contribution to the 
$QQQ$ potential in the different colour representations is what we have called $V_\cc^{(1)}$.
Two classes of diagrams contribute: \emph{two-body diagrams} and
\emph{three-body diagrams}. These are shown in Fig.~\ref{fig:3quark1loop}.

Two-body diagrams are simply the quark-antiquark diagrams of order $g^4$, 
which we know from the static quark-antiquark potential,  
with the static antiquark propagator replaced by a quark propagator and with
the addition of a spectator line. Their colour factor is of course
different but the amplitude can be easily obtained from the
$Q\overline{Q}$ equivalent.

Three-body diagrams such as the ones in Fig.~\ref{fig:3quark1loop} d) 
and \ref{fig:3quark1loop} e) do not contain a spectator quark. 
We will show that diagrams of type  \ref{fig:3quark1loop} d) 
only contribute to the exponentiation of the LO potential, i.e. 
cancel in Eq. (\ref{PTpotential1}) against $ - \mathcal{M}^{(0)\,2}(\cc,\mathfrak{r})/2$,
whereas the ones of type \ref{fig:3quark1loop} e), which include also 
diagrams with two gluons attached to the same quark line,
vanish because they involve triple-gluon vertices of only longitudinal gluons.

\subsection{Calculation of  $V_\cc^{(1)}$}
\label{sub:cooord}
We start by examining the two-body diagrams in Coulomb gauge. 
These are shown in Fig.~\ref{fig:3quark1loop} a), b) (the ladder  and crossed diagrams), c) (the Abelian vertex 
correction) and f) (the gluon self-energy diagrams).
In Coulomb gauge, the crossed diagram and the Abelian vertex vanish. For instance, in position 
space the crossed diagram is proportional to  
\begin{eqnarray*}
&&\int_{-T_W/2}^{T_W/2} \!\!\! dx^0\int_{-T_W/2}^{T_W/2} \!\!\! dx'^0\int_{-T_W/2}^{T_W/2} \!\!\! dy^0
\int_{-T_W/2}^{T_W/2} \!\!\! dy'^0\;\theta(x'^0-x^0)\theta(y^0-y'^0)\delta(x^0-y^0)\delta(x'^0-y'^0)
\\
&=&\int_{-T_W/2}^{T_W/2} \!\!\! dx^0\int_{-T_W/2}^{T_W/2} \!\!\! dx'^0\;\theta(x'^0-x^0)\theta(x^0-x'^0) = 0,
\end{eqnarray*}
where the thetas come from the static quark propagators and the deltas from the longitudinal 
gluon propagators in Coulomb gauge. A similar argument applies to the 
Abelian vertex. In the case of the ladder diagram, the product of deltas and thetas does not yield zero 
but ${T_W^2}/{2}$; the complete result is 
\begin{equation}
\label{ladder3exp}
\mathcal{M}^{(1)}_q(\cc,\br_q)_{\mathrm{lad}}=-g^4f^{(1)}_q(\cc)_{\mathrm{lad}}
\frac{T_W^2}{2}\left[\int\frac{d^3\bq}{(2\pi)^3}\frac{e^{i\bq\br_q}}{\bq^2}\right]^2=
-\als^2f^{(1)}_q(\cc)_{\mathrm{lad}}\frac{T_W^2}{2}\frac{1}{\br_q^2},
\end{equation}
where $f^{(1)}_q(\cc)_{\mathrm{lad}}$ is defined as (we chose $k$ and $k'$ to label the spectator line)
\begin{equation}
\label{colourfactor3q}
f^{(1)}_q(\cc)_{\mathrm{lad}}=\frac{
\cc^u_{ijk} T^a_{ir}T^b_{ri'}T^a_{js}T^b_{sj'} {\cc}_{i'j'k'}^{v\dagger}\delta_{kk'}}{\cc^u_{mno}\cc_{mno}^{v\dagger}}.
\end{equation}
In total, there are three ladder diagram contributions, $\mathcal{M}^{(1)}_q(\cc,\br_q)_{\mathrm{lad}}$, 
with $q$ running from 1 to 3.

Let's consider now the three-body diagram in Fig.~\ref{fig:3quark1loop} d).
We call $\mathcal{M}^{(1)}_{qq'}(\cc,\br_q,\br_{q'})_{3\mathrm{body}}$
its contribution to $\mathcal{M}^{(1)}(\cc,\mathfrak{r})$, which is given by 
\begin{eqnarray}
\nonumber	
\mathcal{M}^{(1)}_{qq'}(\cc,\br_q,\br_{q'})_{3\mathrm{body}}&=&
-g^4f^{(1)}_{qq'}(\cc)_{3\mathrm{body}}\int_{-T_W/2}^{T_W/2} \!\!\! dx^0\int_{-T_W/2}^{T_W/2} \!\!\! dx'^0
\int_{-T_W/2}^{T_W/2} \!\!\! dy^0\int_{-T_W/2}^{T_W/2} \!\!\! dy'^0 \, \theta(y^0-x^0)
\\
\nonumber
&&\times\delta(x'^0-x^0)\delta(y'^0-y^0)
\int\frac{d^3\bq}{(2\pi)^3}\frac{e^{i\bq\cdot\br_q}}{\bq^2}\int\frac{d^3\bq'}{(2\pi)^3}\frac{e^{i\bq'\cdot\br_{q'}}}{\bq'^2}
\\
\label{amplitude3}
&=&-\als^2f^{(1)}_{qq'}(\cc)_{3\mathrm{body}}\frac{T_W^2}{2}\frac{1}{\vert\br_q\vert\,\vert\br_{q'}\vert}.
\end{eqnarray}
In the case of Fig.~\ref{fig:3quark1loop} d),   
the colour factor $f^{(1)}_{qq'}(\cc)_{3\mathrm{body}}$ is defined as
\begin{equation}
\label{colourfactor3b}
f^{(1)}_{qq'}(\cc)_{3\mathrm{body}}=
\frac{\cc^u_{ijk} T^a_{ii'}T^b_{jr}T^a_{rj'}T^b_{kk'}{\cc}_{i'j'k'}^{v\dagger}}{\cc^u_{mno}{\cc}_{mno}^{v\dagger}}.
\end{equation}
In total, there are six three-body diagram contributions, 
$\mathcal{M}^{(1)}_{qq'}(\cc,\br_q,\br_{q'})_{3\mathrm{body}}$, 
with $q$  and $q'$  ($q\neq q'$) running from 1 to 3.

The contributions of the ladder and three-body diagrams cancel in 
Eq. (\ref{PTpotential1}) against 
$\displaystyle - \mathcal{M}^{(0)\,2}(\cc,\mathfrak{r})/2 = 
- \left( \sum_{q=1}^3\mathcal{M}_{q}^{(0)}(\cc,\br_{q})\right)^2/2$.
This happens because 
\begin{eqnarray}
\label{expconditionsquare}
f^{(1)}_q(\cc)_{\mathrm{lad}}=(f^{(0)}_q(\cc))^2\qquad\forall q,
\\
\label{expconditioncross}
f^{(1)}_{qq'}(\cc)_{3\mathrm{body}}=f^{(0)}_q(\cc)f^{(0)}_{q'}(\cc)\qquad\forall q,q'.
\end{eqnarray}
We will prove these identities for all representations $\cc$ in the following 
Sec.~\ref{sub:colourexp}. Hence the ladder and the three-body diagrams 
only contribute to the exponentiation of the LO potential. 

Finally, we are left with the evaluation of the diagram in Fig.~\ref{fig:3quark1loop} f).
The diagram has, in general, a $q^0$ dependence, however, the integration over time 
in the $T_W \to \infty$ limit sets $q^0=0$.\footnote{The  $T_W \to \infty$ limit 
comes from the matching condition \eqref{defpotential2}. It sets $q^0=0$ as in 
$\displaystyle \lim_{T_W\to\infty} \int_{-T_W/2}^{T_W/2} \!\!\!dt \int dq^0 \exp(-iq^0t)\, g(q^0)= g(0)$.
\label{footq0}}
The fermionic part is gauge invariant; 
the gauge part, in Coulomb gauge, may be read, for instance, from \cite{Andrasi:2003zf}.
The one-loop gluon self-energy contribution 
to the gluon propagator in momentum space and at $q^0=0$ is 
\begin{equation}
 \label{dressedgluon} 	
\frac{i\delta^{ab}}{\bq^2}\, \frac{\als}{4\pi}
\left[\left(11-\frac{2}{3}n_f\right)\ln\frac{\mu^2}{\bq^2}
+\frac{31}{3}-\frac{10}{9}n_f\right],
\end{equation}
where $n_f$ is the number of massless light quarks contributing 
to the fermionic part of Fig.~\ref{fig:3quark1loop}~f).
The divergence has been renormalized in the $\MS$ scheme and 
$\mu$ is the renormalization scale.
The contribution to the potential is 
\begin{eqnarray}
V^{(1)}_\cc(\mathfrak{r})&=&\sum_{q=1}^3f^{(0)}_q(\cc) 
\int\frac{d^3\bq}{(2\pi)^3}e^{i\bq\cdot\br_q} \frac{4\pi\als}{\bq^2}\,\frac{\als}{4\pi}
\left[\left(11-\frac{2}{3}n_f\right)\ln\frac{\mu^2}{\bq^2}
+\frac{31}{3}-\frac{10}{9}n_f\right]
\nn
\\
&=& \sum_{q=1}^{3}f^{(0)}_q(\cc)\frac{\als^2}{4\pi\vert\br_q\vert}
\left[2\beta_0(\ln(\mu \vert\br_q\vert) + \gamma_E)+a_1\right],
\label{defpotentialwm1}
\end{eqnarray}
where  $\gamma_E$ is the Euler--Mascheroni constant, 
$a_1=31/3- 10\nf/9$ and $\beta_0=  11 -2\nf/3$.

Since, in Coulomb gauge, all other diagrams of Fig.~\ref{fig:3quark1loop} either vanish or contribute to the 
potential exponentiation, the contribution coming from the diagram in Fig.~\ref{fig:3quark1loop} f)
is the only contribution to the potential at NLO. 
It has the same colour factor as the LO one, which factorizes in front 
of the complete expression of the potential up to NLO.
This reads 
\begin{equation}
\label{1loop3potnomu}
V_\cc(\mathfrak{r})=\sum_{q=1}^{3}f^{(0)}_q(\cc)\frac{\als(1/\vert\br_q\vert)}{\vert\br_q\vert}
\left[1+\frac{\als}{4\pi}\left(2\beta_0\gamma_E + a_1\right)\right],
\end{equation} 
where the colour coefficients $f^{(0)}_q(\cc) $ may be read from Eqs. (\ref{f0singlet}), 
(\ref{f0decuplet}) and (\ref{f0octet}).
We recall that, in the octet case, $V_O$ is a 2$\times$2 matrix. 

The main outcome of Eq. (\ref{1loop3potnomu})
is that at NLO the $QQQ$ static potential and the $Q\bar{Q}$
static potential \cite{Fischler:1977yf} just differ by the overall colour representation, but that the 
effective coupling of the potential, $\displaystyle \alpha_V(1/\vert\br_q\vert) = \als(1/\vert\br_q\vert)
\left[1+\frac{\als}{4\pi}\left(2\beta_0\gamma_E + a_1\right)\right]$, is the same 
for all $Q\bar{Q}$, $QQ$ and $QQQ$ colour representations.
There is no reason to believe that this result keeps holding at NNLO.
Indeed, it has been shown in \cite{Kniehl:2004rk} that the colour-singlet 
and colour-octet effective couplings for the $Q\bar{Q}$ potential differ 
at NNLO.

In Feynman gauge, besides the diagram in Fig. \ref{fig:3quark1loop} f), also 
the diagrams in Fig. \ref{fig:3quark1loop} a), b) and c) contribute 
to the potential. The situation is very similar to the quark-antiquark 
case and it is straightforward to check that the final result up to NLO agrees with 
Eq. (\ref{1loop3potnomu}).

\subsection{Colour factors in the one-loop exponentiation}
\label{sub:colourexp}
In this section, we prove Eqs. \eqref{expconditionsquare} 
and \eqref{expconditioncross} for all colour representations.
In the singlet case, for all $q$ and $q'$ we obtain 
\begin{equation}
f^{(1)}_q(S)_{\mathrm{lad}}= f^{(1)}_{qq'}(S)_{3\mathrm{body}}= \frac{4}{9}.
\label{singlet3quarks1loop}
\end{equation}
Together with Eq. \eqref{f0singlet}, this proves Eqs. \eqref{expconditionsquare} 
and \eqref{expconditioncross}.
Analogously, in the decuplet case, for all $q$ and $q'$ we obtain
\begin{equation}
\label{decupletcheck}
f^{(1)}_q(\Delta)_{\mathrm{lad}}=f^{(1)}_{qq'}(\Delta)_{3\mathrm{body}}=\frac{1}{9}, 
\end{equation}
which again, together with Eq. \eqref{f0decuplet}, proves Eqs. \eqref{expconditionsquare} 
and \eqref{expconditioncross}.
In the octet case, $f^{(1)}_q(O)_{\mathrm{lad}}$, as defined in Eq. \eqref{colourfactor3q}, 
$f^{(1)}_{qq'}(\cc)_{3\mathrm{body}}$, as defined in Eq. \eqref{colourfactor3b}, and 
$f^{(0)}_q(O)$, as defined in Eq. \eqref{f0octet}, are 2$\times$2 matrices.
By explicit computation, one can show that 
\begin{eqnarray}
f^{(1)}_1(O)_{\mathrm{lad}}&=&\left(\begin{matrix}\frac{4}{9}\; &0\\0&\frac{1}{9}\;
\end{matrix}\right) = (f^{(0)}_1(O))^2, 
\quad
f^{(1)}_2(O)_{\mathrm{lad}}= \left(\begin{matrix}\frac{7}{36}&\frac{1}{4\sqrt{3}}\\
\frac{1}{4\sqrt{3}}&\frac{13}{36}\end{matrix}\right) = (f^{(0)}_2(O))^2, 
\nn\\
f^{(1)}_3(O)_{\mathrm{lad}}&=&\left(\begin{matrix}\frac{7}{36}&-\frac{1}{4\sqrt{3}}\\
-\frac{1}{4\sqrt{3}}&\frac{13}{36}\end{matrix}\right) = (f^{(0)}_3(O))^2 ,
\label{squaremi}
\end{eqnarray}
which proves  Eq. \eqref{expconditionsquare}. One can also show that 
\begin{eqnarray}
f^{(1)}_{13}(O)_{3\mathrm{body}} \!\!&=&\!\!\left(\begin{matrix}-\frac{1}{18}&-\frac{1}{2\sqrt{3}}\\
\frac{1}{4\sqrt{3}}&-\frac{5}{36}\end{matrix}\right)
= f^{(0)}_1(O)f^{(0)}_3(O),
\nn\\
f^{(1)}_{12}(O)_{3\mathrm{body}} &=&\left(\begin{matrix}-\frac{1}{18}&\frac{1}{2\sqrt{3}}\\
-\frac{1}{4\sqrt{3}}&-\frac{5}{36}\end{matrix}\right)
= f^{(0)}_1(O)f^{(0)}_2(O),
\label{crossedmi}\\
f^{(1)}_{32}(O)_{3\mathrm{body}} \!\!&=& \!\!\left(\begin{matrix}-\frac{13}{72}&-\frac{\sqrt{3}}{8}\\
\frac{\sqrt{3}}{8}&-\frac{1}{72}\end{matrix}\right)
= f^{(0)}_3(O)f^{(0)}_2(O).
\nn
\end{eqnarray} 
This is enough to prove Eq. \eqref{expconditioncross}, because, $f^{(0)}_q(O)$ being 
symmetric, it holds that$(f^{(0)}_q(O)f^{(0)}_{q'}(O))^T$ $=$ $f^{(0)}_{q'}(O)f^{(0)}_q(O)$, 
and, moreover, $\displaystyle (f^{(1)}_{qq'}(O)_{3\mathrm{body}})^T = f^{(1)}_{q'q}(O)_{3\mathrm{body}}$.\footnote{The 
fact that the $f^{(1)}_{qq'}(O)_{3\mathrm{body}}$ matrices are not
symmetric under the exchange $q\leftrightarrow q'$ does not contradict time-reversal invariance,  
since, in the complete three-body amplitude, for each diagram proportional to 
$f^{(1)}_{qq'}(O)_{3\mathrm{body}}/{(\vert\br_q\vert\,\vert\br_{q'}\vert)}$
there is a diagram proportional to $f^{(1)}_{q'q}(O)_{3\mathrm{body}}/{(\vert\br_q\vert\,\vert\br_{q'}\vert)}$ 
that restores the symmetry.}

\section{The case of  $\nc $ colours and $\nc$  quarks}
\label{sec:N} 
A generalization of Eq. \eqref{qqqtensor} to $\nc>3$ colours but with three
quarks can be easily obtained using the \emph{Hook length formula}
on the corresponding Young tableaux \cite{Fulton}. However, 
a system made of three quarks and $\nc\ne 3$ colours does not 
contain a colour-singlet state. For this reason, in the following, 
we will consider the case of $\nc$ quarks and $\nc$ colours. 
With the increase in the number of quarks, also the number of representations increases rapidly, 
but we will always have a totally antisymmetric representation (the colour-singlet one) 
and a totally symmetric representation, whose dimension is $\displaystyle \binom{2\nc-1}{\nc}$, 
i.e. the number of independent entries in a totally symmetric tensor of
rank $\nc$ with indices running from $1$ to $\nc$.

The singlet representation \eqref{baryonsinglet} can be easily generalized 
to any given number $\nc$ of colours and quarks using the Levi--Civita tensor 
$\varepsilon_{ijk\dots}$ of rank $\nc$. 
Since 
\begin{equation}
\label{epsilonnorm}
\varepsilon_{ijk\dots}\varepsilon_{ijk\dots} = \nc!\,,
\end{equation}
where repeated indices are summed from 1 to $\nc$,  
the normalized totally antisymmetric singlet tensor is given by 
\begin{equation}
\label{baryonsingletnc}
\tilde{S}_{ijkl\ldots}=\frac{\varepsilon_{ijkl\ldots}}{\sqrt{\nc!}},
\end{equation} 
where, from now on, a tilde will designate representations with $\nc$ colours and quarks.
We provide now an expression for the colour factors $f^{(0)}_q(\tilde{S})$ relevant at LO.
Since the singlet tensor is totally antisymmetric, the factors $f^{(0)}_q(\tilde{S})$ are equal for all $q$. 
The product of two Levi--Civita tensors can be expressed as a determinant of Kronecker $\delta$ symbols 
in the following way:
\begin{equation}
\label{epsilonproduct}
\varepsilon_{ijk\dots}\varepsilon_{mnl\dots} = 
\det 
\left(
\begin{matrix}
\delta_{im} & \delta_{in} & \delta_{il} & \dots \\
\delta_{jm} & \delta_{jn} & \delta_{jl} & \dots \\
\delta_{km} & \delta_{kn} & \delta_{kl} & \dots \\
\vdots & \vdots & \vdots \\
\end{matrix}
\right),
\end{equation}
which generalizes the three-dimensional identity 
$\varepsilon_{ijk}\varepsilon_{lmn}=\delta_{il}(\delta_{jm}\delta_{kn}-\delta_{jn}\delta_{km})$ 
$-\delta_{im}(\delta_{jl}\delta_{kn}-\delta_{jn}\delta_{kl})$ 
$+\delta_{in}(\delta_{jl}\delta_{km}-\delta_{jm}\delta_{kl})$. 
Using this property we obtain the colour factor \cite{Cornwall:1996xr}
\begin{equation}
\label{singlet3quarks}
f^{(0)}_q(\tilde{S})=\frac{\varepsilon_{ijkl\ldots}}{\sqrt{\nc!}}T^a_{im}T^a_{jn}\delta_{ko}\delta_{lp}\ldots
\frac{\varepsilon_{mnop\ldots}}{\sqrt{\nc!}}
=-\frac{\cf}{\nc-1}
\quad 
\forall q,
\end{equation} 
where $C_F=(\nc^2-1)/2/\nc$.

The singlet LO potential is then
\begin{equation}
\label{potsinglet}
V^{(0)}_{\tilde{s}}(\mathfrak{r})=-\frac{\cf}{\nc-1}\als\sum_{q=1}^{n}\frac{1}{\vert\br_q\vert},
\end{equation}
where the sum runs over all $n=\nc(\nc-1)/2$ possible one-gluon exchanges between 
two different quark lines and $\mathfrak{r}$ is the dimension $\nc(\nc-1)/2$ vector
$({\bf x}_1-{\bf x}_2, ...,  {\bf x}_1-{\bf x}_N, {\bf x}_2-{\bf x}_3, ..., {\bf x}_{N-1}-{\bf x}_N)$.
We observe that the singlet $Q\bar{Q}$ potential is $\nc -1$ times
each two-body component of the singlet potential of a baryon made of $\nc$ quarks,   
which generalizes the well-known result that the quark-quark
potential in an ordinary baryon ($\nc=3$) is half the quark-antiquark potential.
This may be understood in the following way: if we collapse $\nc-1$ quarks in the same position 
the remaining one will ``see'' $\nc-1$ times the quark-quark potential. This, in turn,  corresponds to the
quark-antiquark potential, since the SU$(\nc)$ antisymmetric representation 
of rank $\nc-1$ describing a system of $\nc-1$ quarks in a totally antisymmetric colour state
has dimension $\nc$ and corresponds to the conjugate of the fundamental
representation, i.e. the representation describing an antiquark.

For what concerns the totally symmetric representation, let $\tilde{\Delta}^u_{ijk\ldots}$ 
be a generic symmetric tensor, with $u$ running from 1 to $\displaystyle \binom{2\nc-1}{\nc}$. 
The totally symmetric equivalent of Eq. \eqref{epsilonproduct} is 
\begin{equation}
\sum_{u=1}^{\binom{2\nc-1}{\nc}}\tilde{\Delta}^u_{ijk\ldots}\tilde{\Delta}^u_{i'j'k'\ldots}=
\frac{1}{\nc!}\sum_{\sigma(i'j'k'\ldots)}\delta_{ii'}\delta_{jj'}\delta_{kk'}\ldots\, ,
\label{projsymmetric}
\end{equation} 
where, on the right-hand side, there are $\nc$ Kronecker deltas and 
the sum is understood to be performed
over all permutations of the indices $i'j'k'\ldots$.
The tensors $\tilde{\Delta}^u_{ijk\ldots}$ are normalized as
\begin{equation}
\tilde{\Delta}^u_{ijkl\ldots}\tilde{\Delta}^v_{ijkl\ldots} = \delta^{uv}\,.
\label{normDeltatilde}
\end{equation}
In analogy with Eq. \eqref{singlet3quarks}, the totally symmetric colour factor 
relevant at LO is 
\begin{equation}
\label{symmetricfactor}
f^{(0)}_q(\tilde{\Delta})=
\frac{
\tilde{\Delta}^u_{ijkl\ldots}
T^a_{ii'}T^a_{jj'}\delta_{kk'}\delta_{ll'}\ldots \tilde{\Delta}^v_{i'j'k'l'\ldots}}
{
\tilde{\Delta}^u_{ijkl\ldots}\tilde{\Delta}^v_{ijkl\ldots}}
=\frac{\cf}{\nc+1}
\quad 
\forall q.
\end{equation}  
The result follows from 
\begin{equation}
T^a_{ii'}T^a_{jj'}=-\frac{\delta_{ii'}\delta_{jj'}}{2\nc}+\frac{\delta_{ij'}\delta_{ji'}}{2},
\label{TaTa}
\end{equation}
the totally symmetric nature of $\tilde{\Delta}^u_{ijkl\ldots}$ and the 
normalization (\ref{normDeltatilde}). The LO totally symmetric potential is then
\begin{equation}
\label{potsymmetric}
V^{(0)}_{\tilde{d}}(\mathfrak{r})=\frac{\cf}{\nc+1}\als\sum_{q=1}^{n}\frac{1}{\vert\br_q\vert},
\end{equation}
where, as before, $n=\nc(\nc-1)/2$.

We prove now the exponentiation of the colour-singlet and colour-symmetric potentials at NLO, 
i.e. Eqs. (\ref{expconditionsquare}) and (\ref{expconditioncross}),  
for a baryon in SU$(\nc)$ made of $\nc$ quarks. 
We can write the colour factor $f^{(1)}_q(\tilde{S})_{\mathrm{lad}}$ as
\begin{displaymath}
 f^{(1)}_q(\tilde{S})_{\mathrm{lad}}=
\frac{\varepsilon_{ijkl\ldots}}{\sqrt{\nc!}}T^a_{ix}T^b_{xm}T^a_{jy}T^b_{yn}
\delta_{ko}\delta_{lp}\ldots\frac{\varepsilon_{mnop\ldots}}{\sqrt{\nc!}},
\end{displaymath} 
and the colour factor $f^{(1)}_q(\tilde{\Delta})_{\mathrm{lad}}$ as 
\begin{displaymath}
f^{(1)}_q(\tilde{\Delta})_{\mathrm{lad}}=
\frac{\tilde{\Delta}^u_{ijkl\ldots}T^a_{ix}T^b_{xm}T^a_{jy}T^b_{yn}
\delta_{ko}\delta_{lp}\ldots\tilde{\Delta}^v_{mnop\ldots}}
{\tilde{\Delta}^u_{ijkl\ldots}\tilde{\Delta}^v_{ijkl\ldots}}.
\end{displaymath} 
Using Eq. (\ref{TaTa}), the totally antisymmetric nature of 
$\varepsilon_{mnop\ldots}$, the totally symmetric nature of 
$\tilde{\Delta}^u_{ijkl\ldots}$ and the normalizations (\ref{epsilonnorm}), (\ref{normDeltatilde})
we obtain
\begin{eqnarray}
\nonumber 
f^{(1)}_q(\tilde{S},\tilde{\Delta})_{\mathrm{lad}} =\left(\frac{\cf}{\nc\mp1}\right)^2
\quad 
\forall q,
\label{singletNcquarks1loopsquare}
\end{eqnarray}
where the upper sign refers to the antisymmetric case and the lower sign to the symmetric one. 
This  proves that $f^{(1)}_q(\tilde{S})_{\mathrm{lad}}$ and $f^{(1)}_q(\tilde{\Delta})_{\mathrm{lad}}$ are 
the squares of $f^{(0)}_q(\tilde{S})$ and $f^{(0)}_q(\tilde{\Delta})$ respectively, 
i.e. Eq. (\ref{expconditionsquare}).

For the three-body diagram we adopt a similar procedure, with the
difference that here the contracted indices will be
$\nc-3$.\footnote{For definiteness, we assume the two gluons to be attached 
to the same quark line. However, starting from $\nc=4$ quark lines, 
it is also possible that a gluon is exchanged between two quarks 
and a second one is exchanged between two different quarks. This is again a
${1}/({\vert\br_q\vert\vert\br_{q'}\vert})$ term and by similar arguments it can be shown that its colour
factor is also the square of \eqref{singlet3quarks}, thus obeying \eqref{expconditioncross}.} 
The colour factors are then, for all $q$ and $q'$,
\begin{displaymath}
f^{(1)}_{qq'}(\tilde{S})_{3\mathrm{body}}=
\frac{\varepsilon_{ijkl\ldots}}{\sqrt{\nc!}}T^a_{ix}T^b_{xm}T^a_{jn}T^b_{ko}\delta_{lp}\ldots
\frac{\varepsilon_{mnop\ldots}}{\sqrt{\nc!}}\,,
\end{displaymath}
and
\begin{displaymath}
f^{(1)}_{qq'}(\tilde{\Delta})_{3\mathrm{body}}=
\frac{\tilde{\Delta}^u_{ijkl\ldots}T^a_{ix}T^b_{xm}T^a_{jn}T^b_{ko}\delta_{lp}\ldots\tilde{\Delta}^v_{mnop\ldots}}
{\tilde{\Delta}^u_{ijkl\ldots}\tilde{\Delta}^v_{ijkl\ldots}}.
\end{displaymath}
Proceeding like before, we obtain 
\begin{equation}
f^{(1)}_{qq'}(\tilde{S},\tilde{\Delta})_{3\mathrm{body}}=\left(\frac{\cf}{\nc\mp1}\right)^2
\quad 
\forall q, q',
\label{singletNcquarks1loopcross}
\end{equation} 
which proves Eq. \eqref{expconditioncross} for the antisymmetric (upper sign) and the symmetric (lower sign) case.

\section{The three-body part of the  static potential at NNLO}
\label{sec:3qg6} 
We may ask when a genuine three-body interaction, i.e. a contribution 
which is not the sum of three ${1}/{\vert\br_q\vert}$ terms
and is not generated by the exponentiation of two-quark interactions, 
shows up in the Wilson loop. This happens at order $g^6$.
More precisely, we write\footnote{We assume that $\ln(\mu |\br_q|)$ terms have been resummed 
such that the potential up to NLO reads
$\displaystyle V_\cc(\mathfrak{r})=\sum_{q=1}^{3}f^{(0)}_q(\cc)\frac{\als(1/\vert\br_q\vert)}{\vert\br_q\vert}
\left[1+\frac{\als(1/\vert\br_q\vert)}{4\pi}\left(2\beta_0\gamma_E + a_1\right)\right]$.
Under this condition, terms like $\ln(\mu |\br_q|)$ or  $\ln^2(\mu |\br_q|)$ are absent at NNLO.} 
\begin{equation}
V^{(2)}_{\mathcal{C}}(\mathfrak{r}) = V^\mathrm{3body}_{\mathcal{C}}(\mathfrak{r}) 
+ \als^3 \sum_{q=1}^3 \frac{a^\mathrm{2body}_{q}(\mathcal{C})}{|{\bf r}_q|},
\label{defthreebody}
\end{equation}
where the three-body part of $V^{(2)}_{\mathcal{C}}$, $V^\mathrm{3body}_{\mathcal{C}}$, 
is defined as the part of $V^{(2)}_{\mathcal{C}}$ that vanishes when putting one of the quarks 
at infinite distance from the other two, i.e. in the limit 
$|{\bf r}_i|$, $|{\bf r}_j|\to\infty$ ($i\neq j$) with fixed $|{\bf r}_k|$ ($k\neq i$ and $k\neq j$). 
Since $V^{(2)}_{\mathcal{C}}$ is gauge invariant, then, by definition, 
also the numerical coefficients $a^\mathrm{2body}_{q}(\mathcal{C})$ 
and $V^\mathrm{3body}_{\mathcal{C}}$ are. $V^\mathrm{3body}_{\mathcal{C}}$ may only stem 
from diagrams with gluons attached to all three quark lines.

\begin{figure}
\begin{center}
\includegraphics{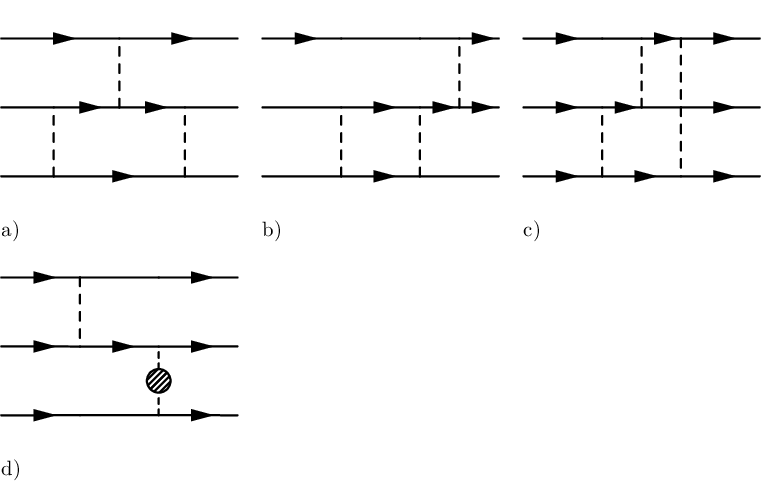}
\end{center}
\caption{Three-body diagrams at NNLO that contribute, in Coulomb gauge, to the exponentiation of the LO and NLO potential.}
\label{fig:exponentiate}
\end{figure}

At order $g^6$, we have many diagrams that involve gluons attached 
to all three quark lines. These can be divided into some basic categories. The adoption of the Coulomb gauge
proves again useful, making only a small subset of these diagrams different from zero. 
We thus have the following diagrams, evaluated, for simplicity, between totally antisymmetric 
and symmetric colour states only.
\begin{enumerate}
\item The diagrams displayed in Fig.~\ref{fig:exponentiate} contribute to the exponentiation 
of the tree-level and one-loop potentials. 
At this order of perturbation theory, the matching condition is given by Eq. \eqref{PTpotential2}. 
It is easily shown that the amplitudes of the diagrams a), b) and c) are\footnote{In the general case 
of a baryon in SU($\nc$) made of $\nc$ quarks,
$$
f^a_{qq'}(\tilde{S},\tilde{\Delta})=f^b_{qq'}(\tilde{S},\tilde{\Delta})=f^c_{qq'q''}(\tilde{S},\tilde{\Delta})
=\left(\mp\frac{\cf}{\nc\mp1}\right)^3,
$$
for all $q$, $q'$ and $q''$. The upper signs refer to the 
antisymmetric (singlet) representation and the lower signs to the symmetric representation.
}
\begin{fmffile}{3qg6exp}
\setlength{\unitlength}{1mm}
\begin{eqnarray}
		\label{aandb}	\parbox{15mm}{
			\begin{fmfchar*}(15,12)
			\fmfstraight
			\fmfleft{g1,qin1,g3,qin2,g5in,qin3,g2}
			\fmf{quark}{qin3,gqq3,qout3}
			\fmf{dashes,tension=0}{gqq1,gqq2} 
			\fmf{dashes,tension=0}{gqq1b,gqq2b} 
			\fmf{plain,tension=2.5}{qin2,gqq2}
			\fmf{phantom}{gqq2,gqq2b}
			\fmf{plain,tension=2.5}{gqq2b,qout2} 
			\fmf{plain,tension=2.5}{qin1,gqq1}
			\fmf{quark}{gqq1,gqq1b}
			\fmf{plain,tension=2.5}{gqq1b,qout1}							  
                        \fmfright{g1out,qout1,g3out,qout2,g5out,qout3,g2out}
			\fmffreeze
			\fmf{quark}{gqq2,int,gqq2b}
			\fmf{dashes,tension=0}{int,gqq3}
\end{fmfchar*}} &=&i\als^3 f^a_{qq'}(\cc)\frac{T_W^3}{3!}\frac{1}{\vert\br_q\vert\vert\br_{q'}\vert\vert\br_q\vert},
\quad	\parbox{15mm}{
			\begin{fmfchar*}(15,12)
				\fmfstraight
				\fmfleft{g1,qin1,g3,qin2,g5in,qin3,g2}
				\fmf{phantom}{qin3,a,b,qout3}
				\fmf{dashes,tension=0}{gqq1,gqq2} 
				\fmf{dashes,tension=0}{gqq1b,gqq2b} 
				\fmf{plain}{qin2,gqq2}
				\fmf{quark}{gqq2,gqq2b}
				\fmf{phantom}{gqq2b,qout2} 
				\fmf{plain}{qin1,gqq1}
				\fmf{quark}{gqq1,gqq1b}
				\fmf{plain}{gqq1b,qout1}							  
                                 \fmfright{g1out,qout1,g3out,qout2,g5out,qout3,g2out}
				\fmffreeze
				\fmf{quark}{gqq2b,int,qout2}
				\fmf{quark}{qin3,a}
				\fmf{plain}{a,b,gqq3}
				\fmf{quark}{gqq3,qout3}
				\fmf{dashes,tension=0}{int,gqq3}
	\end{fmfchar*}} =	i\als^3 f^b_{qq'}(\cc)\frac{T_W^3}{3!}\frac{1}{\vert\br_q\vert^2\vert\br_{q'}\vert},\\
		\label{ci}
		\parbox{15mm}{
		\begin{fmfchar*}(15,12)
			\fmfstraight
			\fmfleft{g1,qin1,g3,qin2,g5in,qin3,g2}
			\fmf{quark}{qin3,3a}
			\fmf{phantom}{3a,3b,3c}
			\fmf{plain}{3a,3b}
			\fmf{quark}{3b,3c}
			\fmf{quark}{3c,qout3}
			\fmf{quark}{qin2,2a}
			\fmf{phantom}{2a,2b,2c}
			\fmf{quark}{2a,2b}
			\fmf{plain}{2b,2c}
			\fmf{quark}{2c,qout2}
			\fmf{quark}{qin1,1a}
			\fmf{phantom}{1a,1b,1c}
			\fmf{quark,tension=.5}{1a,1c}
			\fmf{quark}{1c,qout1}
			\fmf{dashes,tension=0}{1a,2a} 
			\fmf{dashes,tension=0}{2b,3b} 
			\fmf{dashes,tension=0}{3c,1c}
			\fmfright{g1out,qout1,g3out,qout2,g5out,qout3,g2out}
\end{fmfchar*}}&=&i\als^3 f^c_{qq'q''}(\cc)\frac{T_W^3}{3!}\frac{1}{\vert\br_q\vert\vert\br_{q'}\vert\vert\br_{q''}\vert}.
\end{eqnarray}
Keeping in mind that there are three diagrams of the form of
Eq. \eqref{aandb}, one of type a) and two of type b), for each $q,q'$
pair, and six diagrams of the type of Eq. \eqref{ci}, it is easy
to see that their contributions cancel against $- \mathcal{M}^{(0)}(\cc,\mathfrak{r}) 
\mathcal{M}^{(1)}(\cc,\mathfrak{r})$ $+$  $\mathcal{M}^{(0)\,3}(\cc,\mathfrak{r})/3$ 
in the matching condition \eqref{PTpotential2} and therefore do not
contribute to $V^{(2)}_\cc(\mathfrak{r})$. 

The amplitude of diagram d) can be obtained from Eq. \eqref{amplitude3} substituting 
one of the two longitudinal gluon propagators with Eq. \eqref{dressedgluon}, yielding
\begin{equation}
	\label{ampd}
	\parbox{15mm}{
				\begin{fmfchar*}(15,12)
				 					\fmfstraight
				 					\fmfleft{g1,qin1,g3,qin2,g5in,qin3,g2}
				 					\fmf{quark}{qin3,gqq3}
				 					\fmf{plain}{gqq3,int3}
				 					\fmf{quark}{int3,qout3}
				 					\fmf{dashes}{gqq1,loop} 
				 					\fmf{dashes}{loop,gqq2b} 
				 					\fmf{dashes,tension=0}{gqq3,gqq2} 
				 				\fmfv{dec.shape=circle,dec.fill=.5,dec.size=.10w}{loop}	
				 					\fmf{quark,tension=20}{qin2,gqq2,gqq2b,qout2} 
				 					\fmf{quark,tension=20}{qin1,int1}
				 					\fmf{plain,tension=20}{int1,gqq1}
				 					\fmf{quark,tension=20}{gqq1,qout1} 
				 		\fmfright{g1out,qout1,g3out,qout2,g5out,qout3,g2out}
				 		\end{fmfchar*}
			} =	
-\frac{\als^3}{4\pi}\left[2\beta_0(\ln(\mu \vert\br_q\vert) + \gamma_E)+a_1\right] 
f^{(1)}_{qq'}(\cc)_{3\mathrm{body}} \frac{T_W^2}{2}\frac{1}{\vert\br_q\vert\vert\br_{q'}\vert}.
\end{equation} 
Recalling that we already proved the exponentiation
relation of the colour factor $f^{(1)}_{qq'}(\cc)_{3\mathrm{body}}$ in
Sec.~\ref{sub:colourexp}, from Eqs. \eqref{defpotentialwm}  and \eqref{1loop3potnomu} we see that
diagrams of the type d) (two for each $q,q'$ pair) cancel against 
$- \mathcal{M}^{(0)}(\cc,\mathfrak{r}) \mathcal{M}^{(1)}(\cc,\mathfrak{r})$ in Eq. \eqref{PTpotential2}.
\end{fmffile}	
\begin{figure}
\begin{center}
\includegraphics{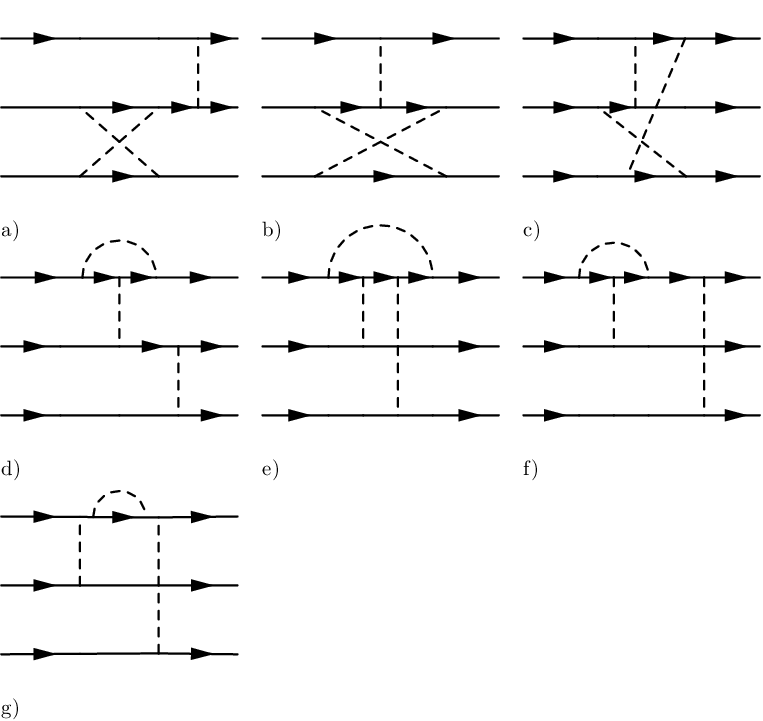}
\end{center}
\caption{Abelian three-body diagrams that have zero amplitude in Coulomb gauge.}
\label{fig:exponentiate_nocoulomb}
\end{figure}
\item Abelian diagrams such as the ones in
Fig.~\ref{fig:exponentiate_nocoulomb} are easily shown
to be zero in Coulomb gauge. However in different gauges, such as the 
Feynman gauge, these diagrams are expected to give a contribution to
the exponentiation and a contribution to the order $\als^3$ result, as
their two-body counterparts do in the $Q\overline{Q}$ case
\cite{Fischler:1977yf,Peter:1997me,Schroder:1998vy}.
\begin{figure}
\begin{center}
\includegraphics{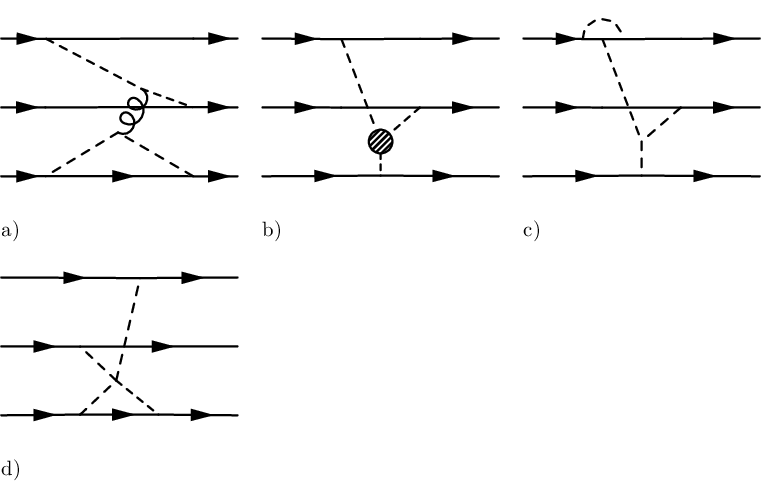}
\end{center}
\caption{Non-Abelian three-body diagrams that have zero amplitude in Coulomb gauge.}
\label{fig:zerocoulomb}
\end{figure}
\item The non-Abelian diagrams shown in Fig.~\ref{fig:zerocoulomb} also vanish.
Diagram a) has a vanishing colour factor between singlet-singlet and decuplet-decuplet initial-final states and, 
in Coulomb gauge, a vanishing amplitude as well.
The dashed blob in diagram b) is a loop of gluons or fermions. 
Lorentz invariance dictates that its Lorentz tensor structure has to be composed by
combinations of a metric tensor $g^{\mu\nu}$ and the external momenta
$q^\lambda_i$. Since the sources are static, this guarantees that the 
Lorentz structure is proportional to at least one power of $q^0_i$.
By means of the usual argument, in the $T_W \to \infty$ limit, 
$q^0_i$ gets multiplied by $\delta(q^0_i)$ and vanishes.
Finally, also diagrams c) and d) vanish 
because they involve non-Abelian vertices with longitudinal gluons only.
\begin{figure}
\begin{center}
\includegraphics[scale=0.8]{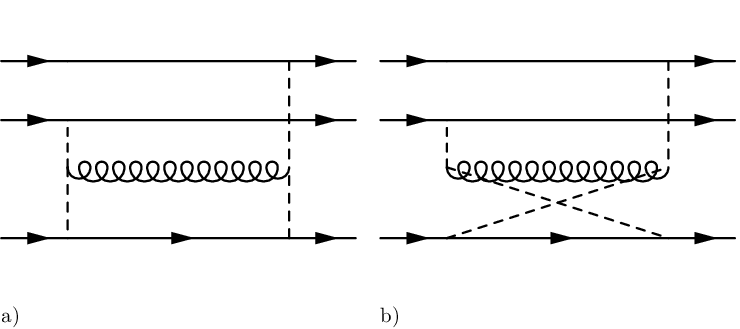}
\end{center}
\caption{The only three-body diagrams that are not exponentiations and that have a non-vanishing amplitude 
in Coulomb gauge. Dashed lines are longitudinal gluons, curly lines are transverse ones.}
\label{fig:3quarkg6}
\end{figure}
\item We are then left with diagrams of the type shown in Fig.~\ref{fig:3quarkg6}:
in Coulomb gauge, these are the only ones contributing to the three-body interaction.
\end{enumerate} 

We now proceed to the evaluation of the diagrams of Fig.~\ref{fig:3quarkg6}. 
There are six different diagrams of each type a) and b): for each source
line there are two diagrams where this line couples to two gluons
(like the bottom line in Fig \ref{fig:3quarkg6}). These two diagrams
are symmetric with respect to a permutation of the other two lines,
 but the independent topologies are just the two shown in Fig.~\ref{fig:3quarkg6}.
We call $\mathcal{H}^a_{\mathcal{C}}$ and $\mathcal{H}^b_{\mathcal{C}}$ 
the momentum-space amplitudes of
the diagrams in Fig.~\ref{fig:3quarkg6} a) and b) respectively and 
$\mathcal{H}_{\mathcal{C}} = \mathcal{H}^a_{\mathcal{C}} + \mathcal{H}^b_{\mathcal{C}}$.

We consider now the colour structure of the diagrams in Fig.~\ref{fig:3quarkg6}.
The colour factors $f_{\mathcal{H}}(S)$ and $f_{\mathcal{H}}(\Delta)$
are equal for all diagrams:
\begin{equation}
\label{fsu33q}
f_{\mathcal{H}}(S)= - \frac{1}{2}\quad \hbox{and} \quad f_{\mathcal{H}}(\Delta)= - \frac{1}{4}.
\end{equation}  
We note that the singlet and decuplet colour factors share the same sign, hence 
also the contributions to the potential from these diagrams will share the same sign, 
at variance with the tree-level and one-loop results.\footnote{For the antisymmetric and symmetric 
representations of a SU$(\nc)$ baryon made of $\nc$ quarks, the colour factors are given by 
\begin{displaymath}
 f_{\mathcal{H}}(\tilde{S})=\frac{\varepsilon_{ijkl\ldots}}{\sqrt{\nc!}}
T^d_{im}T^a_{jn}T^b_{kr}T^e_{ro}f^{bdc}f^{aec}\delta_{lp}\ldots\frac{\varepsilon_{mnop\ldots}}{\sqrt{\nc!}}, 
\end{displaymath}
\begin{displaymath}
f_{\mathcal{H}}(\tilde{\Delta})=
\frac{\tilde{\Delta}^u_{ijkl\ldots}T^d_{im}T^a_{jn}T^b_{kr}T^e_{ro}f^{bdc}f^{aec}
\delta_{lp}\ldots\tilde{\Delta}^v_{mnop\ldots}}
{\tilde{\Delta}^u_{ijkl\ldots} \tilde{\Delta}^v_{mnop\ldots}}.
\end{displaymath}
Using (without summing over $u$),
\begin{eqnarray*}
\frac{\varepsilon_{ijkl_1\ldots l_{N-3}}}{\sqrt{\nc!}} \frac{\varepsilon_{mnol_1\ldots l_{N-3}}}{\sqrt{\nc!}} 
\!\! &=& \!\!
\frac{
  \delta_{im}\left(\delta_{jn}\delta_{ko} - \delta_{jo}\delta_{nk}\right)
- \delta_{in}\left(\delta_{jm}\delta_{ko} - \delta_{jo}\delta_{km}\right)
- \delta_{io}\left(\delta_{jn}\delta_{km} - \delta_{jm}\delta_{kn}\right)}
{N(N-2)(N-1)},
\\
\tilde\Delta^u_{ijkl_1\ldots l_{N-3}} \tilde\Delta^u_{mnol_1\ldots l_{N-3}}
\!\! &=& \!\! 
\frac{
  \delta_{im}\left(\delta_{jn}\delta_{ko} + \delta_{jo}\delta_{nk}\right)
+ \delta_{in}\left(\delta_{jm}\delta_{ko} + \delta_{jo}\delta_{km}\right)
+ \delta_{io}\left(\delta_{jn}\delta_{km} + \delta_{jm}\delta_{kn}\right)}
{N(N+2)(N+1)},
\end{eqnarray*}
we obtain 
$$
f_{\mathcal{H}}(\tilde{S},\tilde{\Delta})= - \frac{\nc\pm 1}{8}.
$$
} 

We compute now $\mathcal{H}_{\mathcal{C}}$. We call ${\bf q}_2$ and ${\bf q}_3$ 
the momenta that flow out of the first and second quark line.  
Setting to zero the external energies, we obtain
\begin{equation}
\label{initialamplitude}
\mathcal{H}^a_{\mathcal{C}}(\bq_2,\bq_3)= - \frac{f_{\mathcal{H}}(\mathcal{C})\,g^6}{\bq_2^2\bq_3^2}
\int\frac{d^4k}{(2\pi)^4}\frac{4
(\bq_2\cdot\hat\bk\,\bq_3\cdot \hat\bk-\bq_2\cdot\bq_3)}
{(k^0+i\epsilon)(\bk-\bq_2)^2(\bk+\bq_3)^2(k^2+i\epsilon)}, 
\end{equation} 
and 
\begin{equation}
\label{initialamplitudeb}
\mathcal{H}^b_{\mathcal{C}}(\bq_2,\bq_3)= - \frac{f_{\mathcal{H}}(\mathcal{C})\,g^6}{\bq_2^2\bq_3^2}
\int\frac{d^4k}{(2\pi)^4}\frac{4
(\bq_2\cdot\hat\bk\,\bq_3\cdot \hat\bk-\bq_2\cdot\bq_3)}
{(-k^0+i\epsilon)(\bk-\bq_2)^2(\bk+\bq_3)^2(k^2+i\epsilon)}.
\end{equation}
Summing $\mathcal{H}^a_\mathcal{C}$ and $\mathcal{H}^b_\mathcal{C}$ yields
\begin{eqnarray}
\mathcal{H}_{\mathcal{C}}(\bq_2,\bq_3)
&=&  - \frac{if_{\mathcal{H}}(\mathcal{C})g^6}{\bq_2^2\bq_3^2}
\int\frac{d^3{\bk}}{(2\pi)^3}\frac{4
(\bq_2\cdot\hat\bk\,\bq_3\cdot \hat\bk-\bq_2\cdot\bq_3)}
{\bk^2(\bk-\bq_2)^2(\bk+\bq_3)^2}
\nonumber\\
&=&
\frac{if_{\mathcal{H}}(\mathcal{C})g^6}{8
\bq_2^2\bq_3^2}
\left[\frac{\vert\bq_2+\bq_3\vert}{\vert\bq_2\vert\vert\bq_3\vert}
+\frac{\bq_2\cdot\bq_3+\vert\bq_2\vert\vert\bq_3\vert}
{\vert\bq_2\vert\vert\bq_3\vert \vert\bq_2+\bq_3\vert}
-\frac{1}{\vert\bq_2\vert}-\frac{1}{\vert\bq_3\vert}\right].
\label{finalamplitude} 
\end{eqnarray}
The contribution of this diagram to the potential in position space is
\begin{equation}
V_{\mathcal{H}\,\mathcal{C}}(\br_2,\br_3)=
i \int \frac{d^3\bq_2}{(2\pi)^3}  e^{i\bq_2\cdot\br_2}  
\int \frac{d^3\bq_3}{(2\pi)^3} e^{i\bq_3\cdot\br_3}
\mathcal{H}_{\mathcal{C}}(\bq_2,\bq_3);
\label{posq2q3}
\end{equation}
the total contribution of all six independent
diagrams of the type 
shown in Fig.~\ref{fig:3quarkg6} is\footnote{%
	The original version of Eq.~\eqref{postot} was off by a factor of 2,
 which is addressed in the current form.
	We are grateful to Beno\^it Assi for pointing this out to us.
	\label{foot_fix}
}  
\begin{equation}
V^\mathrm{tot}_{\mathcal{H}\,\mathcal{C}}(\mathfrak{r})=
V_{\mathcal{H}\,\mathcal{C}}(\br_2,\br_3)
+V_{\mathcal{H}\,\mathcal{C}}(\br_1,-\br_3)
+V_{\mathcal{H}\,\mathcal{C}}(-\br_2,-\br_1).
\label{postot}
\end{equation}

As shown in App.~\ref{app_pos}, $V^\mathrm{tot}_{\mathcal{H}\,\mathcal{C}}(\mathfrak{r})$ 
may be expressed as a double integral suitable for numerical evaluation. 
We have considered the following geometries.

\medskip

\begin{figure}
\begin{center}
\subfigure[]{\includegraphics[scale=0.9]{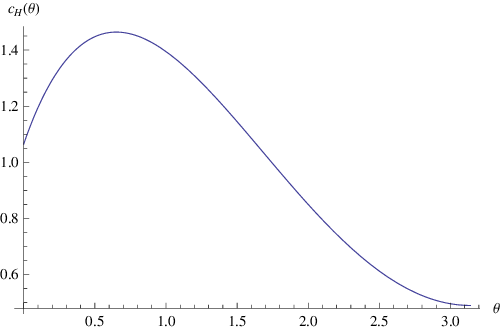}
\label{fig:ch}}
\subfigure[]{
\includegraphics[scale=0.9]{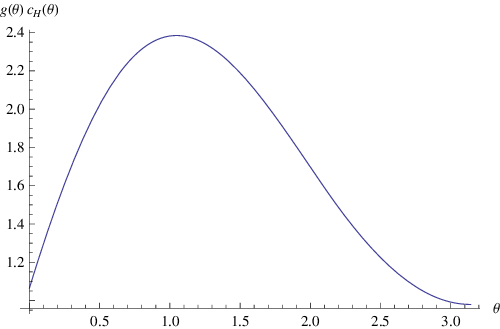}
\label{fig:l}}
\end{center}
\caption{In Fig.~(a), we plot the coefficient $c_{\mathcal{H}}(\theta)$ 
as defined in Eq. \eqref{defch} and obtained from the numerical integration 
of Eq. \eqref{finalposfeyn}. In Fig.~(b), we plot $g(\theta)c_{\mathcal{H}}(\theta)$.}
\label{fig:c}
\end{figure}

{\it (A) Isosceles geometry in a plane} 
\\
In this geometry, the three quarks are placed in different positions of the same 
plane, with two distances chosen to be equal:
$\vert\br_2\vert=\vert\br_3\vert=r$ and $\hat\br_2\cdot \hat\br_3 = \cos\theta$.
The quarks are located at the vertices of an isosceles triangle. 
The potential $V^\mathrm{tot}_{\mathcal{H}\,\mathcal{C}}$ depends on $r$ and $\theta$; 
it has the form\footnote{%
The factor of $1/2$ on the right-hand side of Eq.~\eqref{defch}
keeps track of the change discussed in footnote~\ref{foot_fix}.
}
\begin{equation} 
V^\mathrm{tot}_{\mathcal{H}\,\mathcal{C}}(r,\theta)
=\frac12f_{\mathcal{H}}(\mathcal{C})\als^3 \frac{c_{\mathcal{H}}(\theta)}{r}.
\label{defch}
\end{equation}
In Fig. \ref{fig:ch}, we plot $c_{\mathcal{H}}(\theta)$ as a function of $\theta$. 
The coefficient is always positive, giving rise to an attractive contribution
to the potential, both in the singlet and decuplet channels (we recall that the 
colour factors \eqref{fsu33q} are negative). The dependence on
the angle $\theta$, i.e. on the geometry of the configuration at fixed $r$, is weak: 
$c_{\mathcal{H}}(\theta)$ ranges from a maximum of about $1.46$ 
at $\theta\approx 0.65$ to a minimum of about $0.49$ at $\theta=\pi$. 
On the contrary, the dependence on the geometry of the two-body
contributions to the potential, such as  Eq. \eqref{defpotentialwm}, is
much stronger. In particular, the two-body contribution 
diverges in $|\br_1| = 0$, i.e. for $\theta =0$.

The weaker dependence on the geometry of the three-body contribution 
with respect to the two-body contribution could signal 
the onset of a smooth transition towards the long-distance 
Y-shaped three-body potential seen in the lattice data.
This long-distance potential turns out to depend only on one length, 
$L$, which is the sum of the distances between the
Fermat point of the triangle made of the three quarks and the three quarks.
For isosceles triangles, $L$ has the following dependence on $r$ and $\theta$:
\begin{equation}
L=g(\theta)r,\quad \hbox{where} \quad g(\theta)=
\left\{
\begin{array}{ll}\cos\frac{\theta}{2}+\sqrt{3}\sin\frac{\theta}{2} &\quad \hbox{for} \quad 0\le\theta\le{2\pi}/{3}
\\
2 &\quad \hbox{for} \quad {2\pi}/{3}<\theta\le\pi
\end{array}
\right..
\label{defl}
\end{equation}
Note that the Fermat point of any triangle with an angle greater or equal than $2\pi/3$ 
is located at the vertex of that angle. In terms of $L$, Eq. \eqref{defch} becomes
\begin{equation} 
V^\mathrm{tot}_{\mathcal{H}\,\mathcal{C}}(L,\theta)
=\frac12f_{\mathcal{H}}(\mathcal{C})\als^3 \frac{g(\theta)c_{\mathcal{H}}(\theta)}{L}.
\label{deflh}
\end{equation}
In Fig.~\ref{fig:l}, for completeness, we plot $g(\theta)c_{\mathcal{H}}(\theta)$
as a function of $\theta$. The plot is qualitatively very similar 
to the plot of $c_{\mathcal{H}}(\theta)$: the maximum gets shifted 
to $\theta\approx 1.047$, numerically equivalent to the equilateral geometry $\theta=\pi/3$, 
which thus appears to be the energetically favored one 
for $V^\mathrm{tot}_{\mathcal{H}\,\mathcal{C}}(L,\theta)$ at fixed $L$.

\medskip

{\it (A.1) $\theta=0$: two quarks in the same position} 
\\
A special case of isosceles geometry is $\theta=0$, where two quarks are located in the same position. 
From $\displaystyle \int d(\hat\bq_2\cdot\hat\bq_3)\; \mathcal{H}_{\mathcal{C}}(\bq_2,\bq_3) =0$, 
it follows that $V_{\mathcal{H}\,\mathcal{C}}({\bf 0},\br_3) = V_{\mathcal{H}\,\mathcal{C}}(\br_2,{\bf 0}) =0$, 
hence $V^\mathrm{tot}_{\mathcal{H}\,\mathcal{C}}(r,0) =V_{\mathcal{H}\,\mathcal{C}}(\br,\br)$.
The  three-body potential is finite and given by:
\begin{equation}  
V^\mathrm{tot}_{\mathcal{H}\,\mathcal{C}}(r,0) =\frac12f_{\mathcal{H}}(\mathcal{C})\als^3 \frac{c_{\mathcal{H}}(0)}{r}, 
\quad \hbox{with} \quad  c_{\mathcal{H}}(0)=6-\frac{\pi^2}{2}.
\end{equation}

\medskip

{\it (A.2) $\theta=\pi/3$: planar equilateral geometry} 
\\
In the equilateral case, we have $c_{\mathcal{H}}(\pi/3)\approx 1.377$. 
We may compare the relative magnitude of the three-body contribution to the
tree-level potential. In the singlet case (cf. Eq. \eqref{potsinglet3}), the ratio yields
\begin{equation}
\label{ratio}
\frac{V^\mathrm{tot}_{\mathcal{H}\,s}(r)}{V^{(0)}_s(r)}=
\frac {c_{\mathcal{H}}(\pi/3)}{8
}\als^2(1/r)\approx \frac{\als^2(1/r)}{5.80
},
\end{equation} 
where we have made explicit the scale dependence of the coupling constant.
We note that, using $\als$ at one loop, $V^\mathrm{tot}_{\mathcal{H}\,s}(r)$ 
may become as large as one twelfth
of the tree-level Coulomb potential 
in the region around 0.3 fm, where, at least in the $Q\overline{Q}$ case, 
perturbation theory still holds \cite{Brambilla:2009bi}.

\medskip

{\it (B) Generic geometry} 
\\
In the most general geometry, the three-body potential \eqref{postot} depends on two coordinates.
We may arbitrarily chose one of these coordinates to be $L$, leaving 
the other unspecified. If we call  $a$, $b$, $c$ the lengths of the three sides of the 
triangle made of the three quarks, then $L$ is given by \cite{Takahashi:2000te}
\begin{eqnarray}
\nonumber	
L&=&\left[ \frac{a^2+b^2+c^2}{2} + 
\frac{\sqrt{3(a+b+c)(-a+b+c)(a-b+c)(a+b-c)}}{2}\right]^{\frac12}\!\quad\hbox{for}\quad \theta_{\mathrm{max}}\le\frac{2\pi}{3},
\\
L&=&a+b+c-\mathrm{max}(a,b,c)\quad\hbox{for}\quad\theta_{\mathrm{max}}>\frac{2\pi}{3},
\label{ldef}
\end{eqnarray}
where $\theta_\mathrm{max}$ is the largest angle of the triangle.

\medskip

\begin{figure}
\begin{center}
\includegraphics{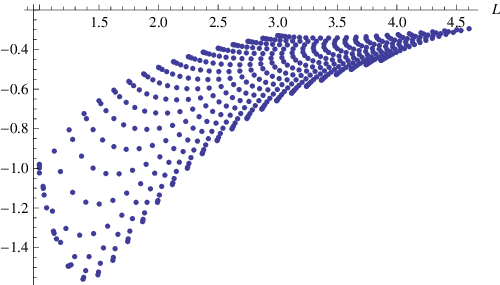}
\end{center}
\caption{The normalized three-body potential, 
$V^\mathrm{tot}_{\mathcal{H}\,\mathcal{C}}(L, ...)/(-f_{\mathcal{H}}(\mathcal{C})\als^3/2)$, 
plotted as function of $L$ for the geometry described in (B.1).}
\label{fig:latticel}
\end{figure}

{\it (B.1) Planar lattice geometry with two fixed quarks} 
\\
In Fig \ref{fig:latticel}, we plot the three-body potential 
obtained by placing the three quarks in a plane $(x,y)$,  
fixing the position of the first quark in  $(0,0)$, the second one  
in $(1,0)$ and moving the third one in the lattice $(0.5 + 0.125\,n_x, 0.125\,n_y)$
with $n_x\in \{0,1, ...,  20\}$ and  $n_y\in \{0,1, ...,  24\}$.
The plot clearly shows the dependence on the geometry at fixed $L$, however, 
the dependence is weaker than in the two-body case.

\medskip

\begin{figure}
\begin{center}
\includegraphics{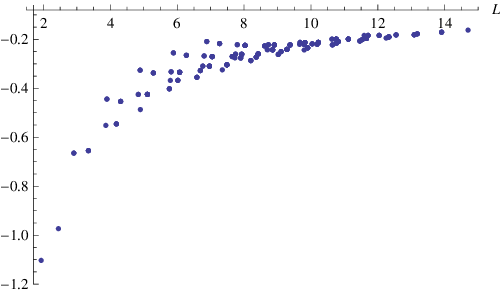}
\end{center}
\caption{The normalized three-body potential, 
$V^\mathrm{tot}_{\mathcal{H}\,\mathcal{C}}(L, ...)/(-f_{\mathcal{H}}(\mathcal{C})\als^3/2)$, 
plotted as function of $L$ for the geometry described in (B.2).}
\label{fig:latticet}
\end{figure}

{\it (B.2) Three-dimensional lattice geometry with the three quarks moving along the axes} 
\\
In the lattice calculation of Ref. \cite{Takahashi:2004rw}, the three quarks 
were located along the axes of a three-dimensional lattice, namely at $(n_x,0,0)$,  $(0,n_y,0)$ and  $(0,0,n_z)$, 
with $n_x \in \{0,1, ..., 6\}$ and $n_y,n_z\in \{1, ..., 6\}$. For the sake of comparison, we consider the same 
geometry and plot the corresponding three-body potential in Fig.~\ref{fig:latticet}.
The plot shows a weak dependence on the geometry: much weaker than in the two-body case, 
but also somewhat weaker than in the geometry considered in (B.1).

\medskip

As a final remark, we would like to note that $V^\mathrm{tot}_{\mathcal{H}\,\mathcal{C}}$, 
the contribution of the diagrams shown in Fig.~\ref{fig:3quarkg6} calculated in Coulomb gauge,  
has an unambiguous physical meaning.
From Eq.~(\ref{finalposfeyn}), it can be seen that this contribution vanishes 
when one of the quarks is put at infinite distance from the other two.
Hence no two-body contribution gets entangled in $V^\mathrm{tot}_{\mathcal{H}\,\mathcal{C}}$, which 
can be rightfully identified with the three-body potential, $V^\mathrm{3body}_{\mathcal{C}}$, defined 
in Eq. \eqref{defthreebody}.

\section{The colour-singlet static potential at NNLO}
\label{sec:nnlo} 
In the colour singlet case, Eq. \eqref{defthreebody} becomes
\begin{equation}
V^{(2)}_{s}(\mathfrak{r}) = V^\mathrm{3body}_{s}(\mathfrak{r}) 
+ \als^3 a^\mathrm{2body}(S) \sum_{q=1}^3 \frac{1}{|{\bf r}_q|}.
\label{defthreebodyS}
\end{equation}
The coefficient $a^\mathrm{2body}(S)$ is independent of the geometry of the 
three quarks: we can take advantage of this fact and calculate $a^\mathrm{2body}(S)$  
without performing any explicit two-loop calculation. 
In a configuration like the one described in (A.1), $V^{(2)}_{s}$ 
is only a function of the distance $r$ between one quark and the other two 
located at the same point:
\begin{equation}
V^{(2)}_{s}(r) = -\left(\frac{3}{2}-\frac{\pi^2}{8
}\right)\frac{\als^3 }{r}+ 2 \als^3 \frac{a^\mathrm{2body}(S)}{r},
\label{defthreebodySiso}
\end{equation}
up to a singular term independent on $r$ that we may drop, for instance, 
by dimensionally regularizing the potential in momentum space.
In this configuration, $V^{(2)}_{s}(r)$ is equal to the static quark-antiquark 
potential, because, when three quarks are in a colour-singlet configuration and 
two of them are located at the same point, these two behave as a an antitriplet 
in colour space, i.e. as an antiquark. 
Owing to the two-loop result of the quark-antiquark potential, we may therefore write \cite{Schroder:1998vy}
\begin{equation}
V^{(2)}_{s}(r) = -\frac{4}{3}\frac{\als^3}{r}\frac{1}{(4\pi)^2} 
\left[ a_2+ \left(\frac{\pi^2}{3}+4\gamma_E^2\right)\beta_0^2+\gamma_E\left(4a_1\beta_0+2\beta_1\right)\right],
\label{deftwobodySiso}
\end{equation}
where $\beta_1 = 102 - 38n_f/3$ and 
\begin{equation}
a_2 = \frac{4343}{18}+36\pi^2-\frac{9}{4}\pi^4+ 66\zeta(3)
-\left(\frac{1229}{27}+\frac{52}{3}\zeta(3)\right)n_f +\frac{100}{81}n_f^2.
\end{equation}
From Eqs. (\ref{defthreebodySiso}) and (\ref{deftwobodySiso}), it follows that
\begin{equation}
a^\mathrm{2body}(S) 
=-\frac{2}{3} \frac{1}{(4\pi)^2} 
\left[ a_2 - 18
\pi^2 + \frac{3\pi^4}{2}
+\left(\frac{\pi^2}{3}+4\gamma_E^2\right)\beta_0^2 + \gamma_E\left(4a_1\beta_0+2\beta_1\right)\right]. 
\label{a2body}
\end{equation}
The complete NNLO expression of the three-quark colour-singlet static potential, 
$V^{(2)}_{s}(\mathfrak{r})$, is then given by Eq. \eqref{defthreebodyS}, where 
$ V^\mathrm{3body}_{s}(\mathfrak{r}) = V^\mathrm{tot}_{\mathcal{H}\,s}(\mathfrak{r})$ can be read from 
Eqs.~\eqref{postot} and \eqref{finalposfeyn}, and $a^\mathrm{2body}(S)$ from Eq.~\eqref{a2body}.
The explicit expression of the colour-singlet static potential up to NNLO is listed in Eq. \eqref{VsNNLO}.

\section{Conclusions}
\label{sec:conc} 
We have studied the static potential of a three-quark system in
perturbation theory up to NNLO. Up to NLO, we have analyzed all the colour
channels (singlet, octets and decuplet) of the SU($3$) case and 
the results have been generalized to SU($\nc$) with $N$ quarks for the
totally antisymmetric and totally symmetric channels.  At LO, the
potential is a sum of three Coulombic one-gluon exchanges between two
of the three quarks.  We have pointed out that, already at this order,
octets mix.  At NLO, after proving the potential exponentiation, the
potential turns out to be simply a sum of two-body contributions,
whose effective coupling $\alpha_V$ is independent of the considered
colour state and is the same as for the $Q\overline{Q}$, $QQ$ and
$QQQ$ potentials. It is expected that $\alpha_V$ becomes dependent on
the colour state at NNLO, as it happens in the $Q\overline{Q}$ case.

At NNLO, the first genuine three-body contribution appears.
Three-body contributions are specific features of the $QQQ$ potential
and for this reason of particular interest.  We have calculated this
contribution, providing numerical results for 
several geometrical confi\-gurations. The general outcome is that 
the dependence on the geometry of the three-body force is weaker than for the two-body force.
Combining the three-body contribution with the two-body contribution extracted 
from the NNLO expression of the quark-antiquark  static potential, we have obtained 
the complete three-quark colour-singlet static potential at NNLO. It reads
\begin{eqnarray}
V_s(\mathfrak{r})&=&-\frac{2}{3}
\sum_{q=1}^{3}\frac{\als(1/\vert\br_q\vert)}{\vert\br_q\vert}
\left\{1+\frac{\als(1/\vert\br_q\vert)}{4\pi}\left[
\frac{31}{3}+22 \gamma_E -\left(\frac{10}{9}+\frac{4}{3}\gamma_E\right)n_f\right]
\right.
\nonumber
\\
&& 
\hspace{20mm}
\left.
+\left(\frac{\als(1/\vert\br_q\vert)}{4\pi}\right)^2
\left[
+66 \zeta(3)+484 \gamma_E^2+ \frac{1976}{3}\gamma_E -
 \frac{3}{4}\pi^4 + \frac{175
}{3}\pi^2 + \frac{4343}{18}
\right.\right.
\nonumber
\\
&& \hspace{50mm}
-\left(\frac{52}{3} \zeta(3) + \frac{176}{3}\gamma_E^2 +\frac{916}{9}\gamma_E 
+ \frac{44}{9}\pi^2 + \frac{1229}{27}\right)n_f 
\nonumber
\\
&& \hspace{50mm}
\left.\left.
+\left(\frac{16}{9}\gamma_E^2 + \frac{80}{27}\gamma_E + \frac{4}{27}\pi^2
+\frac{100}{81}\right)n_f^2
\right]
\right\}
\nonumber
\\
&&
-\als \left(\frac{\als}{4\pi}\right)^2
\left[v_{\mathcal{H}}(\br_2,\br_3)+v_{\mathcal{H}}(\br_1,-\br_3)+v_{\mathcal{H}}(-\br_2,-\br_1)\right],
\label{VsNNLO}
\end{eqnarray}
where
$\displaystyle v_{\mathcal{H}}(\br_2,\br_3) =8
\pi\hat{\br_2}\cdot \hat{\br_3}
\int_0^1dx\int_0^1dy\,\frac{1}{R}
\left[\left(1-\frac{M^2}{R^2}\right)\arctan\frac{R}{M}+\frac{M}{R}\right]$
$\displaystyle + 
8
\pi \hat{\br_2}^i\hat{\br_3}^j$
$\times \displaystyle \int_0^1dx\int_0^1dy\,$ $\displaystyle \frac{\hat{\brg}^i\hat{\brg}^j}{R}
\left[\left(1+3\frac{M^2}{R^2}\right)\arctan\frac{R}{M}-3\frac{M}{R}\right]$,
with 
$\brg=x\br_2-y\br_3$, $R =|\brg|$ and $M=\vert\br_2\vert\sqrt{x(1-x)}+\vert\br_3\vert\sqrt{y(1-y)}$.
Note that by pulling one of the quarks at infinite distance from the others, 
the three-body potential as well as two of the two-body potentials 
vanish and Eq. (\ref{VsNNLO}) reduces to the quark-quark antitriplet static 
potential at NNLO, relevant for $QQq$ baryons.

In \cite{Brambilla:2005yk}, also the three-loop leading logarithmic contribution in the 
infrared cut off has been calculated. Since that calculation does not account 
for the octet mixing, its result applies for geometries where the mixing cancels, 
like the isosceles one. It would be interesting to extend that calculation to 
generic geometries and combine the result with the complete NNLO result given above.

Other possible future developments include comparisons with lattice
results. They exist both for the ground state (the colour-singlet
state) and for the possibly first gluonic excitation of the $QQQ$
system \cite{Takahashi:2002it,Takahashi:2004rw}. An accurate
comparison in the short range will show the running of the three-body
potential and determine at which distances a perturbative description
of the three-body potential breaks down. It may also serve to
establish the nature of the gluonic excitation seen in the lattice
data, determine if it is indeed the first excitation and clarify if,
in the short range, the three static quarks assume a singlet, an octet
or a decuplet colour configuration; it may also serve to extract the
masses of the gluelumps made of three static quarks.  For all this it
is crucial that octet mixing is properly taken into account in the
analysis and in the lattice set up if geometries different from the
isosceles one are used.  Finally, in the case of more general
geometries, it would provide particular insight in the
non-perturbative dynamics of QCD, to investigate the transition region
from (short) distances dominated by two-body forces (where the
potentials depend on two coordinates) to (long) distances dominated by
three-body forces (where, for the Y-shaped configuration, the
potentials depend only on one string length).  In this respect, the
weak dependence on the geometry shown by our results for the leading perturbative 
three-body contribution could indicate a smooth transition to the Y shape.

The $QQQ$ static potential at higher order is relevant for the determination 
of the masses of the baryons made of three heavy quarks. 
Our NLO result is sufficient to provide the masses at NLO,\footnote{If implemented, 
our result may affect the mass determinations obtained  
in Ref. \cite{Jia:2006gw} within a variational study of weakly-coupled baryons.
We note that the value obtained there for the $bbb$ ground state is very 
close to the lattice determination of Ref. \cite{Meinel:2009vv}, providing 
an indirect evidence in support of the Coulombic nature of the system.
}
while at NNLO also $1/m$ and $1/m^2$ potentials should be included.
Clearly, having a reliable determination of the masses is of valuable
help in the experimental searches.

In \cite{Liao:2005hj}, the possible relevance of baryonic states in
the quark-gluon-plasma phenomenology was pointed out and in
\cite{Hubner:2007qh} finite temperature lattice QCD simulations of
$QQQ$ systems in all colour channels were performed.  The lattice data
are very accurate also in the short range and clearly distinguish (in
an equilateral geometry) among the singlet and octet (attractive)
potentials and the (repulsive) decuplet potential before screening
sets in.  Temperature effects at short distances may be systematically
included along the lines developed in Ref. \cite{Brambilla:2008cx} for
the $Q\overline{Q}$ case and comparisons with finite temperature data
may be performed.

In general, one expects that $QQQ$ states in a thermal bath will
experience a much richer phenomenology than $Q\overline{Q}$ states.
First, more colour configurations are possible, second, among these,
not only the singlet but also the octet states are subject, at least
in some geometries, to an attractive interaction. Finally, there will
be a larger variety of possible transitions among the different states
induced by the thermal bath. Thermal transitions between
colour-singlet and colour-octet or colour-decuplet states will likely
be the dominant source of the $QQQ$ colour-singlet thermal decay width
in the short distance, low temperature regime as it is the case for
the colour-singlet to colour-octet transitions in the $Q\overline{Q}$
case \cite{Brambilla:2008cx,Vairo:2009ih}.

\begin{acknowledgments}
We  acknowledge financial support from the RTN Flavianet MRTN-CT-2006-035482 (EU) 
and from the DFG cluster of excellence ``Origin and structure of the universe'' 
(\href{http://www.universe-cluster.de}{www.universe-cluster.de}). 
\end{acknowledgments}

\appendix
\section{Representations}
\label{app_rep}
To ease the reader, we reproduce here from \cite{Brambilla:2005yk} the tensors 
for the singlet, two octet and decuplet representations in which 
the product of three triplet representations of SU$(3)$ may be decomposed.
The totally antisymmetric singlet tensor is
\begin{equation}
\label{baryonsinglet}
S_{ijk}=\frac{\varepsilon_{ijk}}{\sqrt{3!}},
\end{equation}
the octet antisymmetric in the indices $ij$ is
\begin{equation}
\label{oa}
O^{Aa}_{ijk}=\varepsilon_{ijq}T^a_{kq},\qquad O^{Aa*}_{ijk}=\varepsilon_{ijq}T^a_{qk},
\end{equation}
where the index $q$ is summed from $1$ to $3$, the octet symmetric in $ij$ is
\begin{equation}
\label{os}
O^{Sa}_{ijk}=\frac{1}{\sqrt{3}}\left(\varepsilon_{ikq}T^a_{jq}+\varepsilon_{jkq}T^a_{iq}\right),
\qquad 
O^{Sa*}_{ijk}=\frac{1}{\sqrt{3}}\left(\varepsilon_{ikq}T^a_{qj}+\varepsilon_{jkq}T^a_{qi}\right),
\end{equation}
and the symmetric decuplet is
\begin{eqnarray}
\nonumber&&
\hspace{-4mm}
\Delta^1_{111}=\Delta^5_{222}=\Delta^{10}_{333}=1,               \hspace{12mm}
\Delta^2_{112}=\Delta^2_{121}=\Delta^2_{211}=\frac{1}{\sqrt{3}}, \hspace{8mm} 
\Delta^3_{122}=\Delta^3_{221}=\Delta^3_{212}=\frac{1}{\sqrt{3}},\\
\nonumber&&
\hspace{-4mm}
\Delta^4_{113}=\Delta^4_{131}=\Delta^4_{311}=\frac{1}{\sqrt{3}}, \hspace{8mm}
\Delta^7_{133}=\Delta^7_{331}=\Delta^7_{313}=\frac{1}{\sqrt{3}}, \hspace{8mm}
\Delta^8_{223}=\Delta^8_{322}=\Delta^8_{232}=\frac{1}{\sqrt{3}},\\
&&
\hspace{-4mm}
\Delta^9_{233}=\Delta^9_{332}=\Delta^9_{323}=\frac{1}{\sqrt{3}}, \hspace{8mm}
\Delta^6_{123}=\Delta^6_{132}=\Delta^6_{213}=\Delta^6_{231}=\Delta^6_{312}=\Delta^6_{321}=\frac{1}{\sqrt{6}}.
\label{decuplet} 
\end{eqnarray}
One can easily check the following normalization and orthogonality relations:
\begin{eqnarray}
\nonumber&&S_{ijk}S_{ijk}=1,
\quad O^{Aa*}_{ijk}O^{Ab}_{ijk}=O^{Sa*}_{ijk}O^{Sb}_{ijk}=\delta^{ab},
\quad\Delta^\sigma_{ijk}\Delta_{ijk}^\rho=\delta^{\sigma\rho},
\\
&&S_{ijk}O^{Ab}_{ijk}=S_{ijk}O^{Sb}_{ijk}=S_{ijk}\Delta^\sigma_{ijk}=
O^{Aa*}_{ijk}O^{Sb}_{ijk}=O^{Aa*}_{ijk}\Delta^\sigma_{ijk}=O^{Sa*}_{ijk}\Delta^\sigma_{ijk}=0.
\label{normortqqq}
\end{eqnarray}

\section{Position-space three-body potential\label{app_pos}}
From Eqs. \eqref{finalamplitude} and \eqref{posq2q3}, $V_{\mathcal{H}\,\mathcal{C}}(\br_2,\br_3)$ 
may be written as
\begin{eqnarray}
\nonumber
V_{\mathcal{H}\,\mathcal{C}}(\br_2,\br_3)
&=&
f_{\mathcal{H}}(\mathcal{C})g^6\int \frac{d^3\bq_2}{(2\pi)^3}   \int \frac{d^3\bq_3}{(2\pi)^3}
\int\frac{d^3\bk}{(2\pi)^3}\frac{4
(\bq_2\cdot\hat\bk\,\bq_3\cdot \hat\bk-\bq_2\cdot\bq_3) 
e^{i\bq_2\cdot\br_2} e^{i\bq_3\cdot\br_3}}
{\bq_2^2\bq_3^2\bk^2(\bk-\bq_2)^2(\bk+\bq_3)^2}.
\label{q2q3contrib}
\end{eqnarray}
In order to evaluate the integrals, it is convenient to introduce the Feynman parameters $x$ and $y$:
\begin{eqnarray}
\int \frac{d^3\bq_2}{(2\pi)^3}\frac{{\bf q}_2^ie^{i\bq_2\cdot\br_2}}{\bq_2^2(\bk-\bq_2)^2} &=&
-i\partial^i_{\br_2}\int_0^1dx\int \frac{d^3\bq_2}{(2\pi)^3}\frac{e^{i\bq_2\cdot\br_2}}{[\bq_2^2(1-x)+(\bk-\bq_2)^2x]^2},
\label{feynx}
\\
\int \frac{d^3\bq_3}{(2\pi)^3}\frac{{\bf q}_3^ie^{i\bq_3\cdot\br_3}}{\bq_3^2(\bk+\bq_3)^2} &=&
-i\partial^i_{\br_3}\int_0^1dy\int \frac{d^3\bq_3}{(2\pi)^3}\frac{e^{i\bq_3\cdot\br_3}}{[\bq_3^2(1-y)+(\bk+\bq_3)^2y]^2}.
\label{feyny}
\end{eqnarray}
In this form, the integrals in $\bq_2,\bq_3$ and $\bk$ can be performed analytically and 
$V_{\mathcal{H}\,\mathcal{C}}(\br_2,\br_3)$ ends up as a two-dimensional integral in $x$ and $y$:
\begin{eqnarray}
\nonumber 
V_{\mathcal{H}\,\mathcal{C}}(\br_2,\br_3)
&=&\frac{f_{\mathcal{H}}(\mathcal{C})\als^3}{\pi}\, \hat{\br_2}^i\hat{\br_3}^j
\int_0^1dx\int_0^1dy\,\frac{1}{R}\left\{\delta^{ij}
\left[\left(1-\frac{M^2}{R^2}\right)\arctan\frac{R}{M}+\frac{M}{R}\right]\right.
\\
&&
\hspace{40mm}
\left.+ \hat{\brg}^i\hat{\brg}^j
\left[\left(1+3\frac{M^2}{R^2}\right)\arctan\frac{R}{M}-3\frac{M}{R}\right]\right\},
\label{finalposfeyn}
\end{eqnarray}
where $\brg=x\br_2-y\br_3$, $R =|\brg|$ and $M=\vert\br_2\vert\sqrt{x(1-x)}+\vert\br_3\vert\sqrt{y(1-y)}$.

\bibliography{baryQQQ_NLO}

\end{document}